%% file: main_iclr_revised.tex
\documentclass{article}

\usepackage{iclr2027_conference,times}

\usepackage[utf8]{inputenc}
\usepackage[T1]{fontenc}
\usepackage{url}
\usepackage{booktabs}
\usepackage{amsfonts}
\usepackage{amsmath}
\usepackage{amssymb}
\usepackage{bm}
\usepackage{microtype}
\usepackage{graphicx}
\usepackage{subcaption}
\usepackage{enumitem}

\usepackage{multirow}
\usepackage{wrapfig}

\usepackage{algorithm}
\usepackage{algpseudocode}
\usepackage{array}

\usepackage[table,dvipsnames]{xcolor}
\usepackage[colorlinks=true,
            linkcolor=blue,
            citecolor=ForestGreen,
            urlcolor=blue]{hyperref}

\usepackage[most]{tcolorbox}

\definecolor{oursbg}{HTML}{F0F0F0}
\definecolor{gapred}{HTML}{C0392B}

\newcommand{\gap}[1]{\,{\scriptsize\color{gapred}$\downarrow$#1}}
\newcommand{\mtag}[1]{{\scriptsize\color{gray}(#1)}}

\newtcolorbox{taskbox}[2]{
  colback=gray!4,
  colframe=gray!55,
  title=\textbf{#1},
  fonttitle=\small,
  coltitle=black,
  boxrule=0.5pt,
  arc=1.5mm,
  left=1.5mm,
  right=1.5mm,
  top=1mm,
  bottom=1mm,
  before skip=0.6em,
  after skip=0.6em,
  width=0.98\linewidth,
  #2
}
\usepackage{float}
\title{Active Learning for Communication Structure Optimization in LLM-Based Multi-Agent Systems}

\author{
Huchen Yang, Xinghao Dong, Dan Negrut \& Jin-Long Wu \\
University of Wisconsin--Madison \\
\texttt{\{huchen.yang,xdong94,negrut,jinlong.wu\}@wisc.edu}
}

\iclrfinalcopy

\begin{document}

\maketitle
\lhead{Preprint.}

\begin{abstract}
Optimizing the communication structure of large language model based multi-agent systems (LLM-MAS) has been shown to improve downstream performance and reduce token usage. Existing methods typically train on randomly sampled tasks. We show that this optimization is sensitive to the choice of training tasks: downstream accuracy varies widely across random task sets, and adding more random tasks does not consistently improve it. To actively identify the most valuable tasks from a pool of task-answer pairs, we propose an ensemble-based information-theoretic task selection framework. It estimates task informativeness by how much a candidate task changes the distribution over graph parameters, using ensemble Kalman inversion as an efficient and derivative-free approximation of the Bayesian update. The resulting estimator is especially suitable for black-box and noisy multi-agent systems and remains effective with ensembles much smaller than the parameter dimension. To further reduce the number of tasks whose informativeness must be evaluated, we introduce representative pre-selection and surrogate modeling. We validate the framework in benign and adversarial settings and across multiple task formats. It outperforms random training, demonstrating more effective task selection and greater overall cost efficiency.

\end{abstract}

\section{Introduction}\label{sec: introduction}

Large language model (LLM) based agents are now applied beyond generic reasoning to more complex problems, such as clinical decision support~\citep{li2024agent, wang2025survey} and scientific machine learning~\citep{wang2025chronollm, gaonkar2025sciml}. LLM-based multi-agent systems (LLM-MAS) further extend this direction by coordinating multiple specialized agents through inter-agent communication~\citep{li2024survey,he2025llm}. In many existing systems, however, communication structures are empirically predefined, such as fully connected or heuristically designed interaction graphs, rather than optimized for the target task distribution~\citep{hong2023metagpt, qian2024chatdev}. Such empirical designs of communication structures have negative impacts on downstream performance of LLM-MAS, e.g., efficiency, accuracy, stability, and robustness. Specifically, redundant edges can waste token usage, and unrestricted communication can propagate hallucinations or adversarial content~\citep{zeng2025s2, shen2025understanding}. Therefore, communication-structure optimization is an important problem in LLM-MAS~\citep{zhang2024aflow,yang2025topological,li2026assemble, zhang2025multi}. By removing redundant interactions and restricting harmful communication pathways, an optimized communication structure can improve the overall downstream performance of LLM-MAS, while also helping identify less reliable agents and limit their influence~\citep{zhang2024cut,leong2025amas,zhang2024g, wang2025agentdropout, bi2025optagent}. 


We observe that communication-structure optimization of LLM-MAS can be highly sensitive to the choice of training tasks. Existing methods typically select training tasks randomly~\citep{zhang2024cut, leong2025amas}, even though different tasks may favor substantially different communication structures~\citep{zhang2024g,zhang2025multi}. We show in Section~\ref{sec: motivation} that communication-structure optimization under random task selection is unreliable in two respects: downstream accuracy varies substantially across independent runs, and increasing the number of training tasks does not consistently improve accuracy, with accuracy decreasing after reaching its peak. These observations suggest that the key question is not only how many tasks are used for training, but also which tasks are used. This naturally motivates an active learning perspective: can we identify the tasks that are most informative for communication-structure optimization?

The key challenge in this active learning perspective is to estimate task informativeness efficiently and reliably for communication-structure optimization. This is a three-level black-box estimation problem: candidate tasks are evaluated through stochastic and non-differentiable LLM-MAS rollouts; structure optimization is performed on top of this process, using RL or policy-gradient methods~\citep{bi2025optagent,zhang2025multi}; and task utility (i.e., information gain) is further defined by the induced change in the communication-structure parameters. This nesting makes task-utility estimation computationally expensive. Direct sample-based utility estimators typically require many rollouts or repeated optimization runs, while gradient- or sensitivity-based estimators, when applied to this nested black-box pipeline, also require repeated rollouts to control variance. We therefore adopt an ensemble-based information-theoretic approach, which avoids end-to-end gradient estimation, reduces the need for large-scale repeated sampling, and enables more efficient and reliable task selection.

In this work, we first empirically characterize the sensitivity of structure optimization to training-task selection. To address this issue, we propose an ensemble-based information-theoretic active learning framework. We quantify task usefulness through the information gain in the graph parameters, using ensemble Kalman inversion (EKI)~\citep{iglesias2013ensemble, kovachki2019ensemble} to efficiently approximate Bayesian posterior updates. This EKI-based gain favors learnable tasks~\citep{mindermann2022prioritized}, i.e., neither too hard to learn nor already learned. We further introduce representative pre-selection and surrogate modeling to reduce the number of tasks whose information gain must be evaluated. We validate the framework in benign and adversarial settings, on tasks ranging from knowledge and reasoning to code generation, multi-step tool use, and long-horizon embodied interaction. To assess the efficiency of EKI, we also examine how the ensemble size affects task selection and downstream accuracy, including ensembles much smaller than the parameter dimension. Our key contributions are summarized as follows:

\begin{itemize}
    \item First, we show that communication-structure optimization for LLM-MAS is sensitive to the choice of training tasks: randomly selected tasks can lead to substantially degraded downstream accuracy. To our knowledge, we are also the first to show that active task selection can mitigate this problem.
    \item Second, we propose an ensemble-based, derivative-free, and information-theoretic active learning approach that estimates information gain with EKI. Under the same selection budget, it is among the best-performing informativeness-based criteria we compare, while some common criteria perform close to random selection.
    \item Third, to our knowledge, we are the first to show that EKI-based task selection remains effective even when the ensemble size is much smaller than the parameter dimension (6 particles for 36 parameters, 10 for 100, and 15 for 400). 
\end{itemize}

\section{Empirical motivation: sensitivity to training-task selection}\label{sec: motivation}

\begin{wrapfigure}[14]{r}{0.33\linewidth}
    \centering
    \vspace{-1.5em}
    \includegraphics[width=0.9\linewidth]{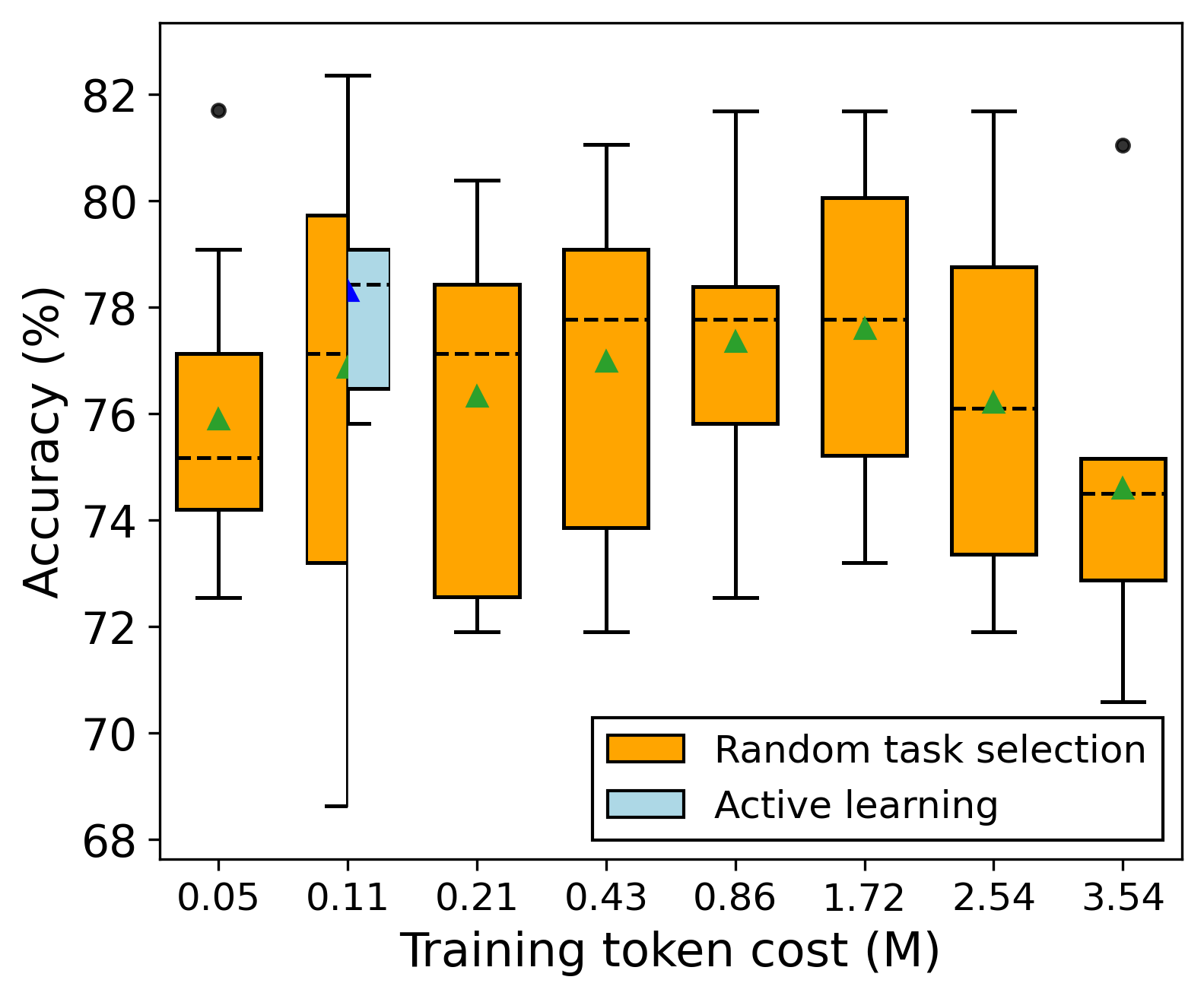}
    \caption{Accuracy-cost scaling on MMLU (adversarial). More randomly selected training tasks help only modestly, while active learning achieves higher accuracy despite lower cost.}
    \label{fig: scaling}
\end{wrapfigure}
Why does task selection matter? We first observe that when the communication structure is optimized using a randomly sampled set of training tasks, the resulting downstream accuracy exhibits substantial variation across independent runs. This instability indicates that communication-structure optimization can be sensitive to the particular training tasks selected.

A natural way to reduce this instability is to increase the number of training tasks. Figure~\ref{fig: scaling} examines this strategy over an extended range of training budgets. Increasing the random-training budget initially provides some improvement, but the gain remains limited and does not consistently increase with additional cost. For reference, our active-learning configuration uses 0.80M tokens for task selection and 0.12M for graph training, resulting in a total cost of 0.92M tokens. Notably, random training still underperforms active task selection even when its training cost substantially exceeds the total selection-and-training cost of active learning. Thus, over the evaluated range, simply spending more computation on randomly selected training tasks does not close the performance gap with active selection.

Accordingly, we formulate task selection as an active learning problem on top of communication-structure optimization and develop a method for identifying informative training tasks.



\section{Related work}\label{sec: related work}

Task selection in this work generally belongs to active learning, but it differs in whether labels are known~\citep{mindermann2022prioritized, yang2025active}. Standard active learning usually decides which data to label and then uses the labeled data for training~\citep{settles2009active}. In our problem, the task--answer pairs are already known, so the selection directly decides which tasks to use for training. Therefore, the cost we aim to reduce comes from training rather than from acquiring labels. The overall problem formulation is still the same.

Active learning strategies often rely on diversity-based, uncertainty-based, or information-theoretic criteria, with practical methods often combining these perspectives~\citep{tharwat2023survey}. Among them, information-theoretic approaches are particularly appealing because they aim to select tasks according to the information they provide about model parameters~\citep{kruschke2008bayesian}. Recent work has proposed many new criteria~\citep{xia2024less, zhang2025d3, lin2025lead, chen2025mig}, but many of them still follow the information-theoretic paradigm~\citep{guilhoto2026mutual, durmazkeser2026large, huang2025aline}. However, estimating information gain is often substantially more difficult than measuring diversity or predictive uncertainty~\citep{kirsch2019batchbald, ren2021survey}.

Information-gain estimation can be approached in two broad ways. The first compares the probability distributions over parameters before and after observing a candidate task, while the second measures how much a deterministic quantity, such as a point estimate or a gradient-based update, would change. The distributional perspective usually requires approximating the Bayesian update induced by the candidate task, using Monte Carlo or other sample-based estimators~\citep{settles2008active, siddhant2018deep, li2026expected, callahanreverse}. Although these sample-based estimators expose uncertainty-related signals, they are often prohibitively expensive~\citep{houlsby2011bayesian, gal2017deep}, especially in LLM-MAS, where scoring each candidate may require repeated multi-agent rollouts across multiple samples or parameter realizations. A cheaper alternative is to rely on gradient-based proxies for parameter updates~\citep{cohn1994improving, cai2013maximizing, xia2024less}. In LLM-MAS, although such gradients may be approximated, the resulting estimators are often noisy and have large variance, making them a less reliable surrogate for information-theoretic utility. This creates a gap for practical active learning methods that efficiently approximate information gain while retaining uncertainty-related signals in black-box multi-agent settings.

\begin{figure}[H]
    \centering
    \includegraphics[width=\linewidth]{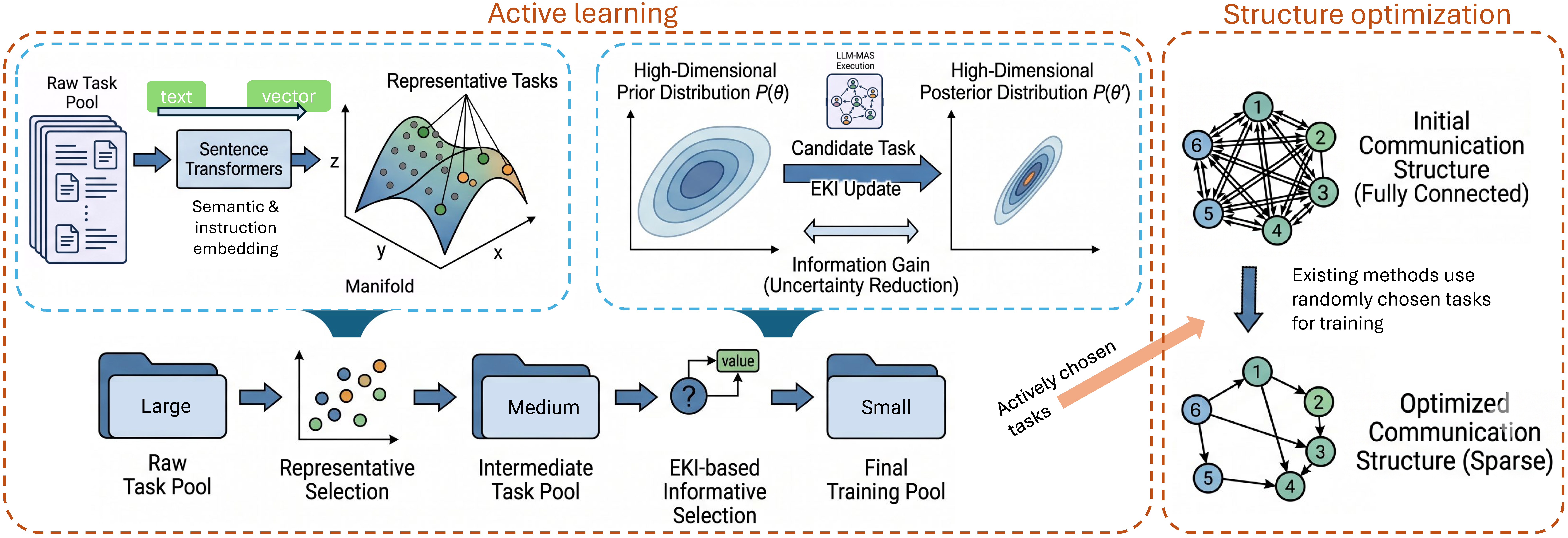}
    \caption{Overview of the proposed active learning framework. The framework first selects representative tasks from a large raw task pool, and then identifies the most informative tasks for communication-structure optimization via one-step EKI-based utility estimation. The selected tasks are finally used to optimize the communication structure.}
    \label{fig: schematic}
\end{figure}

\section{Methodology}\label{sec: methodology}

\subsection{Problem setup}\label{sec: Problem Setup}

\textbf{Communication graph.} We represent the MAS communication structure as a spatial-temporal graph and introduce trainable variables that parameterize edge activation. Specifically, we define $\mathcal{G}=(V,E^S,E^T)$, where \(V=\{v_1,\dots,v_N\}\) is the set of agents, and \(E^S,E^T\in\{0,1\}^{N\times N}\) are the spatial and temporal adjacency matrices describing communication within and across interaction rounds, respectively. An entry \(E_{ij}=1\) indicates that communication from agent \(i\) to agent \(j\) is allowed, while \(E_{ij}=0\) means that the corresponding edge is absent and cannot be activated.

Following a typical setup~\citep{zhang2024cut}, we introduce trainable graph logits as the graph parameters to be optimized:
\[
\mathbf{z}
=
\{Z^S,Z^T\},
\qquad
Z^S,Z^T\in\mathbb{R}^{N\times N}.
\]
Before each MAS rollout, a stochastic mask \(A=\{A^S,A^T\}\) is sampled, with each entry following a Bernoulli distribution whose probability is given by the sigmoid of the corresponding logit. The mask is then applied element-wise to \(E^S\) and \(E^T\), and the masked communication structure is used for the MAS rollout.

\textbf{MAS rollout and forward map.} Given a task \(q\) and graph logits \(\mathbf{z}\), the MAS executes a rollout over the masked communication graph. Each agent receives the task together with messages from its spatial and temporal in-neighbors, and the system produces a final output \(g\) after several rounds of interaction. We represent the entire rollout as
\begin{equation}
    g = G(q;\mathbf{z},\omega),
\end{equation}
where \(g\) is the MAS output, such as a multiple-choice answer, a numerical value, or generated code. The variable \(\omega\) represents the stochasticity in the rollout, for example, from LLM sampling, multi-agent interaction, and final decision.

\textbf{Communication-structure optimization (hereafter, graph optimization).} Given a training dataset $\mathcal{D}_{\mathrm{tr}}=\{(q_i,y_i)\}_{i=1}^M$, where $q_i$ is the task to be solved and $y_i$ is the ground truth or its abstract representation, let $l=\phi(g,q_i,y_i)$ denote the MAS performance score (e.g., answer correctness or code execution success) under the graph logits $\mathbf{z}$, computed by the evaluator $\phi(\cdot)$. The graph optimization problem is to find the graph logits that maximize the expected task performance score over the mask distribution:
\begin{equation}
    \mathbf{z}^\star
=
\arg\max_{\mathbf{z}}
\;
\mathbb{E}_{A\mid \mathbf z}
\mathbb{E}_{(q_i,y_i)\in\mathcal{D}_{\mathrm{tr}}}
\big\{\phi(g,q_i,y_i)\big\},
\label{eq: graph optimization}
\end{equation}
where $\mathbb{E}_{A\mid\mathbf z}$ denotes the expectation over the stochastic masks induced by
the logits $\mathbf z$. All other MAS components, including agent roles, prompts, tools, and the LLM backbone, remain fixed during optimization. The graph logits are optimized with REINFORCE-estimated gradients~\citep{zhang2024cut, leong2025amas, wang2025agentdropout}, though other methods are applicable.

\textbf{Active learning}. In principle, one may optimize \(\mathbf{z}\) using all available training tasks. However, in LLM-MAS, graph optimization is expensive because it requires repeated LLM agent inference. We therefore seek to select a small subset $\mathcal{S}\subsetneq \mathcal{D}_{\mathrm{tr}}$, containing the tasks that are most informative for graph optimization, with the goal of maximizing downstream performance under a limited training budget. Given a task budget $B\ll M$, we formulate the selection problem as
\begin{equation}
    \mathcal{S}^\star
=
\arg\max_{\mathcal{S}\subsetneq \mathcal{D}_{\mathrm{tr}},|\mathcal{S}|\le B\, }
U(\mathcal{S}),
\end{equation}
where the utility function \(U(\mathcal{S})\) quantifies the usefulness of selected tasks for graph optimization.

\textbf{Utility function}. We measure the usefulness of a task set $\mathcal{S}$ by the information gain $I(\mathcal{S})$ on the graph parameters $\mathbf{z}$ (logits). It compares the prior $p(\mathbf{z})$ with the posterior $p(\mathbf{z}|\mathcal{S})$ after a Bayesian update on $\mathcal{S}$. This posterior-to-prior change is a standard information-theoretic measure of Bayesian information gain, widely used in Bayesian experimental design~\citep{chaloner1995bayesian} and Bayesian active learning~\citep{mackay1992information}. The most common metric is the Kullback--Leibler divergence:
\begin{equation}
U(\mathcal{S}) = I(\mathcal{S}) =  D_\mathrm{KL} (p(\mathbf{z}|\mathcal{S})||p(\mathbf{z})),
\label{eq: utility function}
\end{equation}
Other distributional metrics can also be applied, e.g., Wasserstein distances~\citep{helin2025bayesian, yang2026bayesian}. A larger $I(\mathcal{S})$ indicates a more substantial update to the graph parameters and thus greater parameter-level informativeness for graph optimization. Task difficulty, agent disagreement, or prediction diversity do not directly quantify this parameter-change information. Potential failure modes and corresponding mitigation strategies are discussed in Appendix~\ref{apd:failure_modes}.

This formulation is related to expected information gain (EIG) in Bayesian experimental design, which is equivalent to BALD in Bayesian active learning~\citep{houlsby2011bayesian}. In standard EIG, the observation associated with a candidate experiment is unknown before the experiment is performed, so the information gain is averaged over possible observations. In our setting, the analogous observation is the reference answer to a candidate task, which is already available in the training pool. Therefore, no expectation over possible answers is required, and we directly evaluate the realized information gain. For unlabeled or generated tasks whose reference answers are unknown, an outer expectation over possible answers would recover the standard EIG or BALD.

The original graph optimization problem in Eq.~\eqref{eq: graph optimization} is defined on the deterministic variable $\mathbf{z}$ and can be solved by deterministic methods. However, to evaluate information gain by distribution difference in Eq.~\eqref{eq: utility function}, $\mathbf{z}$ should be treated as a random variable and an update process from prior to posterior distribution of $\mathbf{z}$ is required.

\subsection{EKI-Based Task Utility Estimation}\label{sec: eki}

Standard methods for Bayesian updates from prior $p(\mathbf{z})$ to posterior $p(\mathbf{z}|\mathcal{S})$, such as MCMC and SMC, are expensive and computationally impractical for MAS. We therefore efficiently approximate the Bayesian update with ensemble Kalman inversion (EKI,~\citep{iglesias2013ensemble, kovachki2019ensemble}). For notation and pedagogical clarity, we first formulate this update on one task at a time, that is, to estimate $p(\mathbf{z}|q,y)$ and $I(q)$. We then extend it to a multi-task setting for the batch active learning setup to estimate $I(\mathcal{S})$ in Section~\ref{sec: batch}. 

\textbf{Ensemble Kalman inversion.} In this work, EKI serves as an efficient approximation to the Bayesian update in Eq.~\eqref{eq: utility function} and has a deep connection to score-based posterior sampling~\citep{song2020score,chung2022diffusion}: 
\begin{equation}
    \nabla_\mathbf{z} \log p(\mathbf{z}|q,y) = \nabla_\mathbf{z} \log p(\mathbf{z}) +\nabla_\mathbf{z} \log p(y|\mathbf{z},q),
    \label{eq: posterior score}
\end{equation}
where $\nabla_\mathbf{z} \log p(\mathbf{z}|q,y)$ is the score function of the posterior distribution that enables sampling methods such as classical Langevin dynamics~\citep{roberts1996exponential,song2019generative}, or more recent score-based diffusion models~\citep{song2020score,chung2022diffusion}. Some more details of the connection between EKI and score-based posterior sampling are presented in Appendix~\ref{apd:eki_score_da_connection}.

More specifically, EKI represents the uncertainty over the graph parameters using an ensemble of particles $\{\mathbf{z}^{(j)}\}_{j=1}^{J} \in \mathbb{R}^{d_z}$, which are subsequently updated using the task $q$ and answer $y$. 

For each ensemble particle \(\mathbf{z}^{(j)}\), we perform a forward evaluation: we run the LLM-MAS on a given task $q$, obtain the output $g^{(j)}$, and compare it with the true answer $y$ through evaluator $\phi(\cdot)$ to obtain the corresponding scalar performance score $l^{(j)} \in \mathbb{R}$ (Note this score is not restricted to be a scalar; see the multi-objective extension in Appendix~\ref{apd: multi obj}):
\begin{equation}
    g^{(j)} = G(q;\mathbf{z}^{(j)}, \omega), \quad l^{(j)}=\phi(g^{(j)},q,y).
\end{equation}
Together, these forward evaluations form the EKI prediction step. Let $\bar{\mathbf{z}} = \frac{1}{J}\sum_{j=1}^J \mathbf{z}^{(j)},\, \bar{l} = \frac{1}{J}\sum_{j=1}^J l^{(j)}$ denote the empirical means of the parameter ensemble and the predicted scores. The empirical cross-covariance and score covariance are then estimated as
\begin{equation}
\begin{aligned}
    C^{zl}
=
\frac{1}{J-1}\sum_{j=1}^J
\bigl(\mathbf{z}^{(j)}-\bar{\mathbf{z}}\bigr)
\bigl(l^{(j)}-\bar{l}\bigr)^\top, \quad C^{ll}
=
\frac{1}{J-1}\sum_{j=1}^J
\bigl(l^{(j)}-\bar{l}\bigr)
\bigl(l^{(j)}-\bar{l}\bigr)^\top.
\end{aligned}
\label{eq:Czg}
\end{equation}

In the update step, EKI updates each ensemble particle by
\begin{equation}
\mathbf{z}^{(j)}_{\mathrm{post}}
=
\mathbf{z}^{(j)}
+
C^{zl}
\bigl(C^{ll}+\Gamma\bigr)^{-1}
\bigl(l^*-l^{(j)}\bigr),
\label{eq:eki_update}
\end{equation}
where $l^*$ is the desired value for the performance score, e.g., 1 for binary correctness. $\Gamma$ is the score noise covariance, which is a representation of the forward stochasticity $\omega$. In practice, we use it primarily to regularize the update for numerical stability, rather than as a precisely calibrated noise covariance. The updated ensemble $\{\mathbf{z}^{(j)}_{\mathrm{post}}\}_{j=1}^J$ provides an empirical approximation to the posterior distribution \(p(\mathbf{z}|q,y)\)~\citep{yang2025bayesian}. To save computational cost, we only perform a one-step EKI update in this work, which has been shown to be effective for distinguishing different tasks~\citep{yang2025bayesian, callahanreverse}; we further validate this choice in Appendix~\ref{apd:sensitivity_to_step}.

\textbf{Ensemble-based utility approximation.} Given the initial and updated ensembles, the information gain of a single task is approximated by
\begin{equation}
        I(q) = D_\mathrm{KL} (p(\mathbf{z}|q,y)||p(\mathbf{z}))
         \approx D_\mathrm{KL}(\{\mathbf{z}^{(j)}_{\mathrm{post}}\}_{j=1}^J||\{\mathbf{z}^{(j)}\}_{j=1}^{J}).
    \label{eq: u ensemble}
\end{equation}
In practice, we employ Gaussian approximations to the ensembles and compute the KL using their empirical means and variances. We then rank candidate tasks by their estimated information gain.

\subsection{Remarks on the EKI-based utility}

\textbf{Exactness.} The EKI update equals the Bayesian update only under linear-Gaussian assumptions. We therefore use the EKI-based KL only as a low-cost surrogate, not as an accurate estimate of the information gain. We demonstrate its effectiveness for task selection throughout Section~\ref{sec: experiment}.

\textbf{Learnable tasks.} Eq.~\eqref{eq:eki_update} also shows which tasks are informative. If all $J$ ensemble members answer a task correctly, or all answer it wrongly, then $l^{(j)}$ is the same for every $j$. In that case $C^{zl}=0$, the ensemble does not move, and the information gain is zero. Only tasks that some members solve and others fail can move the ensemble. We call these tasks learnable \citep{mindermann2022prioritized}.

\textbf{Role in training.} The EKI update is used only for utility estimation. After task selection, the selected tasks are used to optimize the same graph parameters with REINFORCE.


\subsection{Extension to batch setting}\label{sec: batch}

\textbf{Batch formula}. The single-task formula can be extended to batch setting for the batch gain $I(\mathcal{S})$~\citep{kirsch2019batchbald}. Given an ensemble particle $\mathbf{z}^{(j)}$ and a set $\mathcal{S}=\{(q_i,y_i)\}_{i=1}^m \subseteq \mathcal{D}_{\mathrm{tr}}$ with multiple tasks and answers, the corresponding scores $\{l^{(j)}_i=\phi(g^{(j)}_i,q_i,y_i) \in \mathbb{R}\}_{i=1}^m$ for all tasks can be obtained. To keep the updating formula consistent with Eqs.~\eqref{eq:Czg} and~\eqref{eq:eki_update}, we replace the scalar score with a vector score $\mathbf{l}^{(j)}=[l^{(j)}_1, l^{(j)}_2, \cdots, l^{(j)}_m]^\top \in \mathbb{R}^m$. The corresponding updated ensemble $\{\mathbf{z}^{(j)}_{\mathrm{post}}\}_{j=1}^J$ then represents the posterior updated on all the tasks in $\mathcal{S}$ and the batch information gain $I(\mathcal{S})$ is obtained.


\textbf{Cost of batch evaluation}. For a batch $\mathcal{S}$ of size $m$, evaluating $I(\mathcal{S})$ requires $Jm$ forward evaluations, since each task is evaluated for each ensemble particle. This is of the same order as computing the individual gains $I(q)$ for all tasks in the batch separately. The batch formulation only changes the dimension of the update step, where the covariance and update are computed from the vector-valued scores. In practice, this additional update cost is negligible compared with the forward evaluations.


\textbf{Reuse strategy}. A reuse strategy reduces the total forward evaluations when evaluating overlapping candidate subsets under the batch setting. For example, consider two candidate subsets $\mathcal{S}_1=\{q_1,q_2\}$ and $\mathcal{S}_2=\{q_2,q_3\}$. If the two subset gains are evaluated independently, the prediction step requires $4J$ forward evaluations. However, \emph{if both subsets are evaluated from the same initial ensemble}, the prediction results for the shared task $q_2$ can be reused. Thus, only the three unique tasks $\{q_1,q_2,q_3\}$ need to be evaluated, giving $3J$ forward evaluations. More generally, for a collection of candidate subsets $\{\mathcal{S}_b  \subseteq \mathcal{D}_{\mathrm{tr}}\}_{b=1}^B$, the prediction-step cost is reduced from $J\sum_{b=1}^B |\mathcal{S}_b|$ to $J\left|\bigcup_{b=1}^B \mathcal{S}_b\right|$.

After the scores of each task are cached, each subset gain is obtained by selecting the corresponding cached scores for the tasks within this subset, stacking them into a batch-level vector, and applying the EKI update step. This reuse strategy avoids repeated forward evaluations for the same task whenever it appears in multiple candidate subsets, which can substantially reduce the dominant cost of batch utility estimation. Appendix~\ref{apd:batch_reuse} provides the detailed formulation and discussion.

\subsection{Active learning procedure}\label{sec: al}

Given a way to quantify the information gain, we can enumerate all candidate tasks, evaluate their information gain, and choose either the top-$k$ most informative tasks or form a coreset to optimize the graph. However, for a large candidate pool, enumeration is expensive. To reduce the token cost, we introduce a two-stage selection procedure to shrink the candidate pool for information-gain-based selection, and further introduce surrogate modeling to reduce the number of forward evaluations for information gain. These components can be removed: Information-gain-based selection alone still selects good tasks (ablation in Appendix~\ref{apd:ablation}) only at higher cost. An overall schematic figure is shown in Fig.~\ref{fig: schematic} and the algorithm is given in Appendix~\ref{apd: algorithm}.

We focus on a single-round setting, where task selection and subsequent graph optimization are each performed once, without alternating. A multi-round setting can be achieved by repeating this procedure with the updated graph, but is not considered here.



\textbf{Two-stage selection.} From a large pool, we first select the most representative tasks to form a smaller intermediate pool, from which we then choose the most informative tasks according to their information gain (we call this informative selection). This design allows the inexpensive representative selection to screen a large task pool, while the more expensive informative selection is performed only on a much smaller set. On this reduced pool, either sequential decision-making or enumeration becomes feasible. For extremely large or even infinite datasets, we first randomly sample a candidate pool of manageable size and then apply the same two-stage selection procedure.

\textbf{Representative selection.} We embed the tasks $q_i$ from text into vectors $v_i$ such that their relative positions in the dataset manifold can be identified. The most intuitive approach is to directly apply a text encoder based on sentence transformers~\citep{reimers-2019-sentence-bert, wang2020minilm}. Recent work has proposed instruction embeddings to enhance this process~\citep{li2024instruction}. For tasks with labels that distinguish them into several groups within the whole dataset, e.g., subjects in MMLU~\citep{hendryckstest2021}, these labels can also be included in the embedding process to specify different manifolds for each group. For tasks without explicit labels, e.g., GSM8K~\citep{cobbe2021gsm8k}, we simply input the whole task into a single encoder without further modification.

Once the embedding vectors are obtained, we employ a greedy approach to form the intermediate pool. We first choose the center point, and then sequentially choose the farthest point from the selected points until the budget is reached.

\textbf{Surrogate modeling for informative selection.} To avoid enumerating all candidates in the representative pool, we apply a surrogate modeling approach. We denote the map from a task embedding vector $v(q)$ to its information gain $U(q)$ as $\psi(\cdot)$, $\psi: v(q) \mapsto U(q)$. Specifically, for a given task $q$, we perform EKI to update prior knowledge of the graph $\mathbf{z}$ to a posterior and quantify their difference as the information gain $U(q)$. After several pairs $\{q_k,U_k\}_{k=1}^K$ are collected, with $K$ much smaller than the pool size, we build a surrogate model $\bar{\psi}(\cdot)$ to approximate the true map $\psi(\cdot)$. We use the surrogate model to predict the information gain of unevaluated tasks and choose the most informative tasks for actual evaluation (see details in Appendix~\ref{apd: ts}). These evaluations in turn help improve the surrogate model. This surrogate-based sequential strategy substantially reduces the number of tasks requiring actual evaluation. 

\section{Experiments}\label{sec: experiment}

\subsection{Experimental setup}

We ask whether active task selection improves communication-structure optimization over random selection. We first evaluate this on general-knowledge and mathematical-reasoning tasks under both benign and adversarial settings, where we compare the two methods under both matched training budgets and comparable total budgets to assess performance and cost efficiency. We then extend the evaluation to code generation, multi-step tool use, and long-horizon embodied interaction to examine generalization across broader benchmarks.

In the benign setting, all agents follow their assigned roles; in the adversarial setting, a subset is instructed to provide incorrect responses and mislead the other agents, and no agent is informed whether any other agent is adversarial (Appendix~\ref{apd:mas_setup}). We evaluate downstream performance and token cost, reporting the mean, run-to-run standard deviation, and lower-tail statistics (Q1 and worst-25\% mean) across independent runs. More experimental details can be found in Appendix~\ref{apd: details}.

\subsection{Performance on knowledge and reasoning benchmarks}\label{sec: results}

We first evaluate active task selection on MMLU and GSM8K under both benign and adversarial settings. Both benchmarks use six GPT-4.1-nano agents; in the adversarial setting, three of the six agents are adversarial. Additional experiments varying the number of agents and the LLM backbone are reported in Appendix~\ref{apd:scale_backbone}.

\begin{wrapfigure}[12]{r}{0.25\linewidth}
    \centering
    \vspace{-1.5em}
    \includegraphics[width=\linewidth]{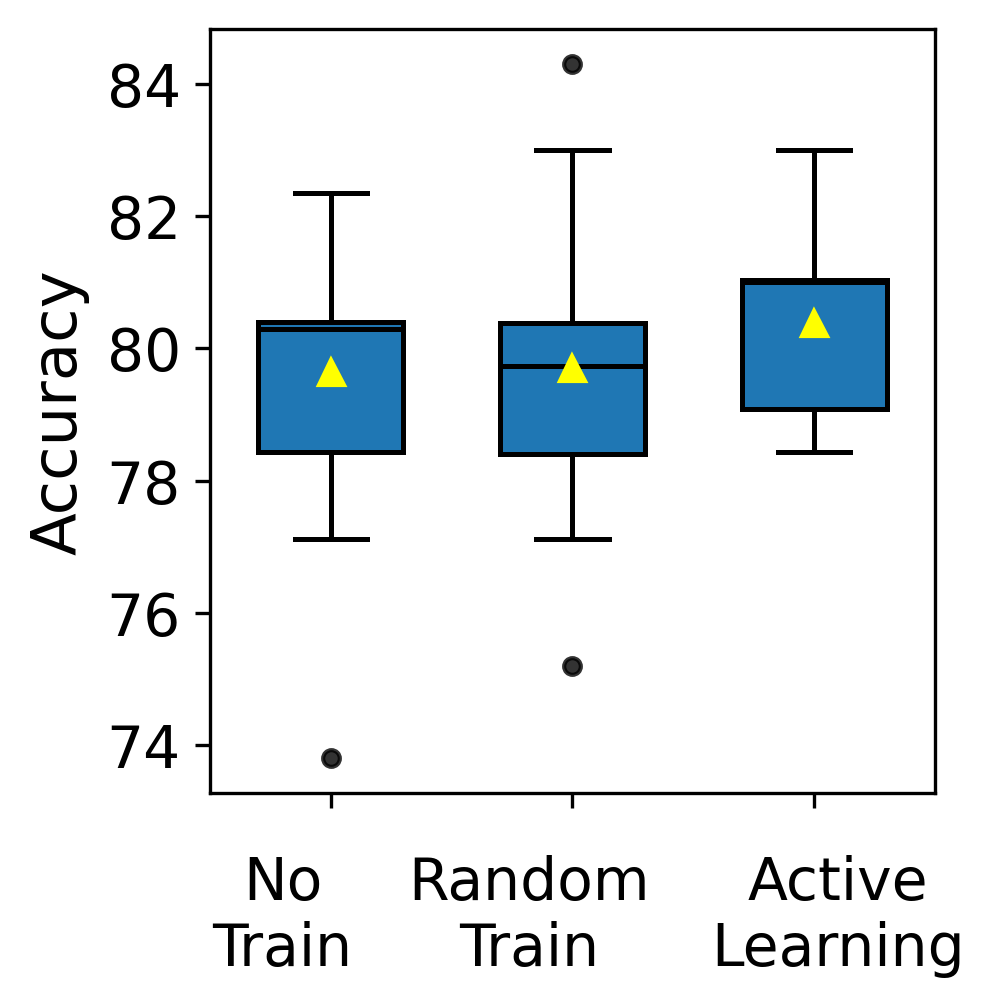}
    \caption{Distribution of downstream accuracy on MMLU (benign)}
    \label{fig:dis_mmlu}
    \vspace{-0.5em}
\end{wrapfigure}

Under the same training budget, active selection improves mean accuracy, run-to-run stability, and lower-tail performance over random selection in all four settings (Table~\ref{tab:main_results}; statistical tests in Appendix~\ref{apd:stat_summary}). A representative run-to-run accuracy distribution is shown in Fig.~\ref{fig:dis_mmlu}, showing less poor runs. Together, they indicate that the selected tasks are more effective for communication-structure optimization and therefore lead to better downstream performance.

The improvement is not simply obtained by paying an additional selection cost: under a comparable total budget, active selection still outperforms random training on both adversarial benchmarks. From the complementary perspective, active selection can achieve a given level of downstream performance with a smaller total budget than random training, further demonstrating its overall cost efficiency.

\begin{table}[h]
\caption{Performance comparison on MMLU and GSM8K under benign and adversarial settings. Accuracy is reported as mean $\pm$ std, first quartile (Q1), and mean below Q1. Random (match training cost only) uses the same token budget as the training step of Active. Random (match total cost) uses a budget comparable to the selection and training steps of Active combined. }
\label{tab:main_results}
\centering
\small
\begin{tabular}{cllll}
\toprule
& & \multicolumn{3}{c}{Accuracy (\%)} \\
\cmidrule(lr){3-5}
Dataset & Setup & Mean $\pm$ std & Q1 & Worst-25\% \\
\midrule
  & No train & 79.64 $\pm$ 1.96 & 78.43 & 77.23 \\
  & Random \mtag{match training cost only} & 79.69 $\pm$ 2.19\gap{0.69} & 78.40\gap{0.68} & 77.26\gap{1.30} \\
\rowcolor{oursbg}
\cellcolor{white}\multirow{-3}{*}{\shortstack{MMLU\\(Benign)}}
  & Active   & \textbf{80.38} $\pm$ 1.35 & \textbf{79.08} & \textbf{78.56} \\
\midrule
  & No train & 92.27 $\pm$ 0.54 & 91.73 & 91.61 \\
  & Random \mtag{match training cost only} & 92.78 $\pm$ 0.54\gap{0.29} & 92.49\gap{0.33} & 92.06\gap{0.71} \\
\rowcolor{oursbg}
\cellcolor{white}\multirow{-3}{*}{\shortstack{GSM8K\\(Benign)}}
  & Active   & \textbf{93.07} $\pm$ 0.28 & \textbf{92.82} & \textbf{92.77} \\
\midrule
  & No train & 75.68 $\pm$ 2.10 & 75.16 & 73.41 \\
  & Random \mtag{match training cost only} & 76.87 $\pm$ 3.72\gap{1.45} & 73.20\gap{3.27} & 72.00\gap{4.25} \\
  & Random \mtag{match total cost} & 77.35 $\pm$ 2.64\gap{0.97} & 75.82\gap{0.65} & 73.85\gap{2.40} \\
\rowcolor{oursbg}
\cellcolor{white}\multirow{-4}{*}{\shortstack{MMLU\\(Adversarial)}}
  & Active   & \textbf{78.32} $\pm$ 1.80 & \textbf{76.47} & \textbf{76.25} \\
\midrule
  & No train & 90.97 $\pm$ 1.32 & 89.83 & 89.56 \\
  & Random \mtag{match training cost only} & 92.90 $\pm$ 1.00\gap{0.93} & 92.17\gap{0.98} & 91.69\gap{1.24} \\
  & Random \mtag{match total cost} & 92.64 $\pm$ 0.71\gap{1.19} & 92.11\gap{1.04} & 91.73\gap{1.20} \\
\rowcolor{oursbg}
\cellcolor{white}\multirow{-4}{*}{\shortstack{GSM8K\\(Adversarial)}}
  & Active   & \textbf{93.83} $\pm$ 0.76 & \textbf{93.15} & \textbf{92.93} \\
\bottomrule
\end{tabular}
\end{table}

\subsection{Generalization to broader agentic benchmarks}

To evaluate whether active task selection generalizes beyond single-response QA, we further consider HumanEval for code generation~\citep{chen2021codex}, InterCode-SQL for multi-step tool use~\citep{yang2023intercode}, and ALFWorld for long-horizon embodied interaction~\citep{ALFWorld20}. All three experiments use adversarial agent configurations, with different numbers of total agents (5/5/4) and adversarial agents (3/1/2) across benchmarks.

For the two multi-step benchmarks, Success@$k$ measures the fraction of tasks solved within $k$ interaction steps, while AUC@$k$ averages success over the interaction horizon and therefore also reflects how early a task is solved.

\begin{figure}[t]
    \centering
    \begin{minipage}[t]{0.48\linewidth}
        \centering
        \includegraphics[width=\linewidth]{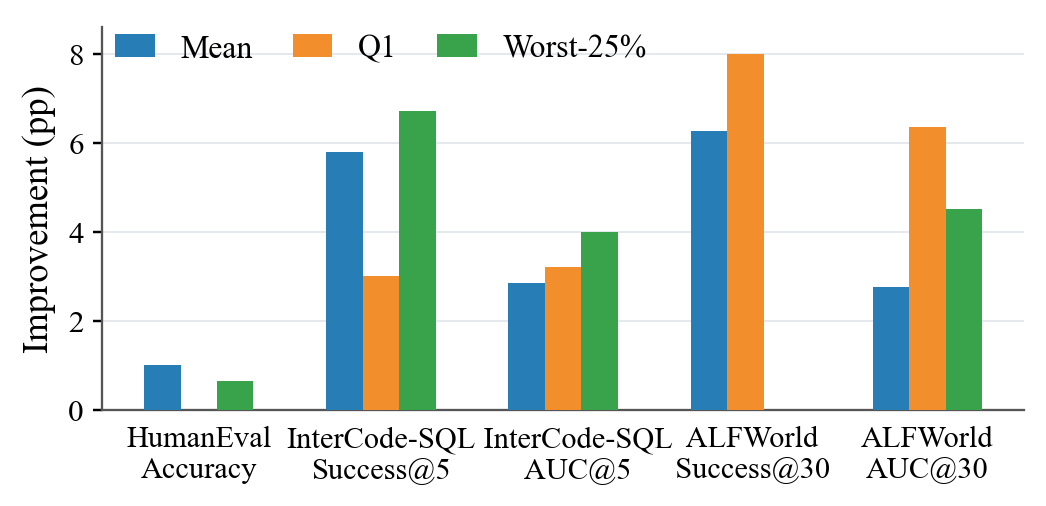}
        \captionof{figure}{Improvement of active task selection over random training across benchmarks with different task formats.}
        \label{fig:broader_benchmarks}
    \end{minipage}
    \hfill
    \begin{minipage}[t]{0.48\linewidth}
        \centering
        \includegraphics[width=\linewidth]{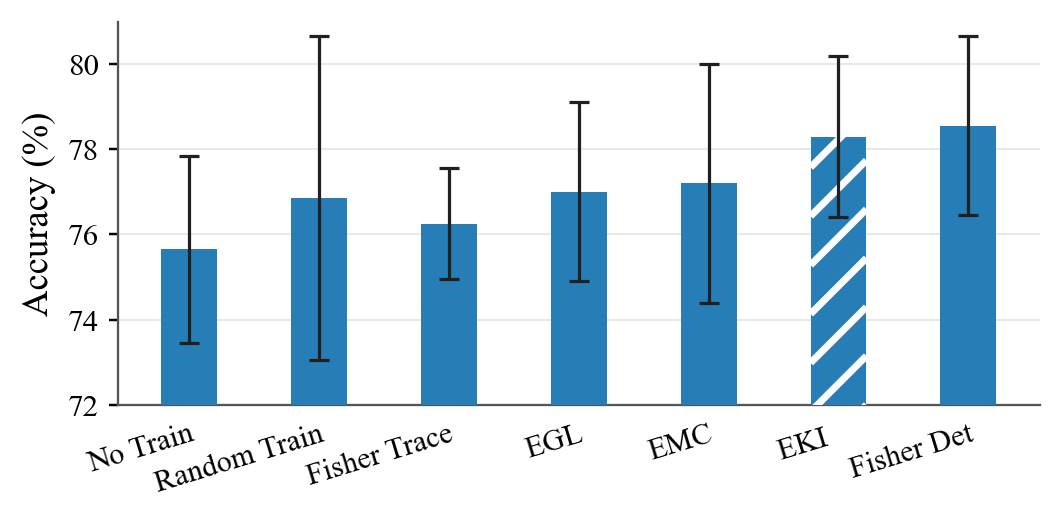}
        \captionof{figure}{Comparison among methods. EKI and Fisher coreset (det) perform best overall, while EKI does not use a pool-level coreset.}
        \label{fig:other_al_methods}
    \end{minipage}
\end{figure}

Across all three benchmarks, active task selection shows improvement over random training on nearly all reported metrics (Fig.~\ref{fig:broader_benchmarks}; Appendix~\ref{apd:broader_benchmark_statistics}). Taken together, these results provide evidence that our active task-selection method remains effective across substantially different task formats and agent–environment interaction forms. More broadly, our method naturally accommodates diverse task formats because the ensemble-based formulation treats MAS rollouts as black-box evaluations.

\subsection{Comparison with other informativeness-based active learning methods}

We compare our method with several informativeness-based active learning baselines on MMLU under agent attacks. All methods share the same 1000-to-50 representative-selection stage and differ only in how they select 10 tasks from the 50-task pool.

We consider expected gradient length (EGL) and expected model change (EMC), which quantify informativeness through the magnitude of parameter changes induced by a task. We also include Fisher-information-based coreset-style baseline with both trace and determinant criteria, which are related to A-optimal and D-optimal experimental designs, respectively. See details in Appendix~\ref{apd: other methods}.


Under the same selection cost, EKI and Fisher coreset (det) perform best among the compared criteria (Fig.~\ref{fig:other_al_methods}), and they are not statistically distinguishable (Appendix~\ref{apd:detail_statistics_other_al}). The choice between them depends mainly on whether gradients can be estimated accurately and on the available memory (Appendix~\ref{apd:eki_fisher_when}). We also compare their selection behavior in Appendix~\ref{apd:eki_fisher_examples}.



\subsection{EKI with small ensembles}\label{sec:small_ensemble}

\begin{figure}[h]
    \centering
    \begin{minipage}[t]{0.48\linewidth}
        \centering
        \includegraphics[width=\linewidth]{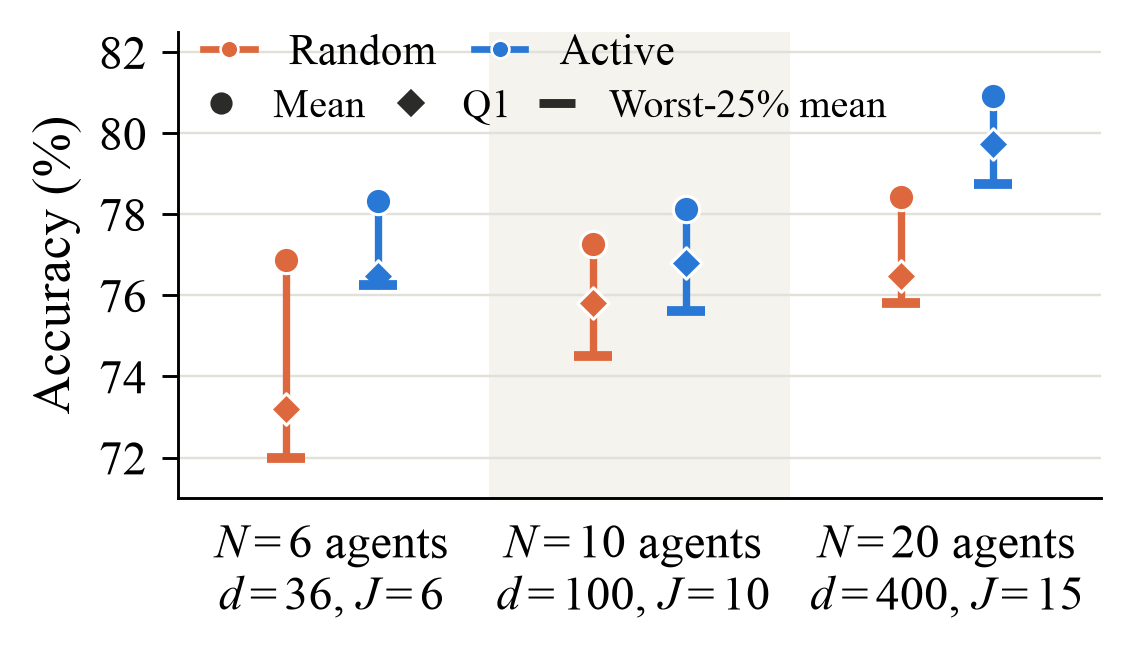}
        \caption{\textbf{Consistent accuracy improvement on larger systems:} with $N$ agents, $d$ parameters, and $J \ll d$ ensemble members for task selection. Details in Appendix~\ref{apd:scalebili}.}
\label{fig:accuracy_small_ens}
    \end{minipage}
    \hfill
    \begin{minipage}[t]{0.48\linewidth}
        \centering
        \includegraphics[width=\linewidth]{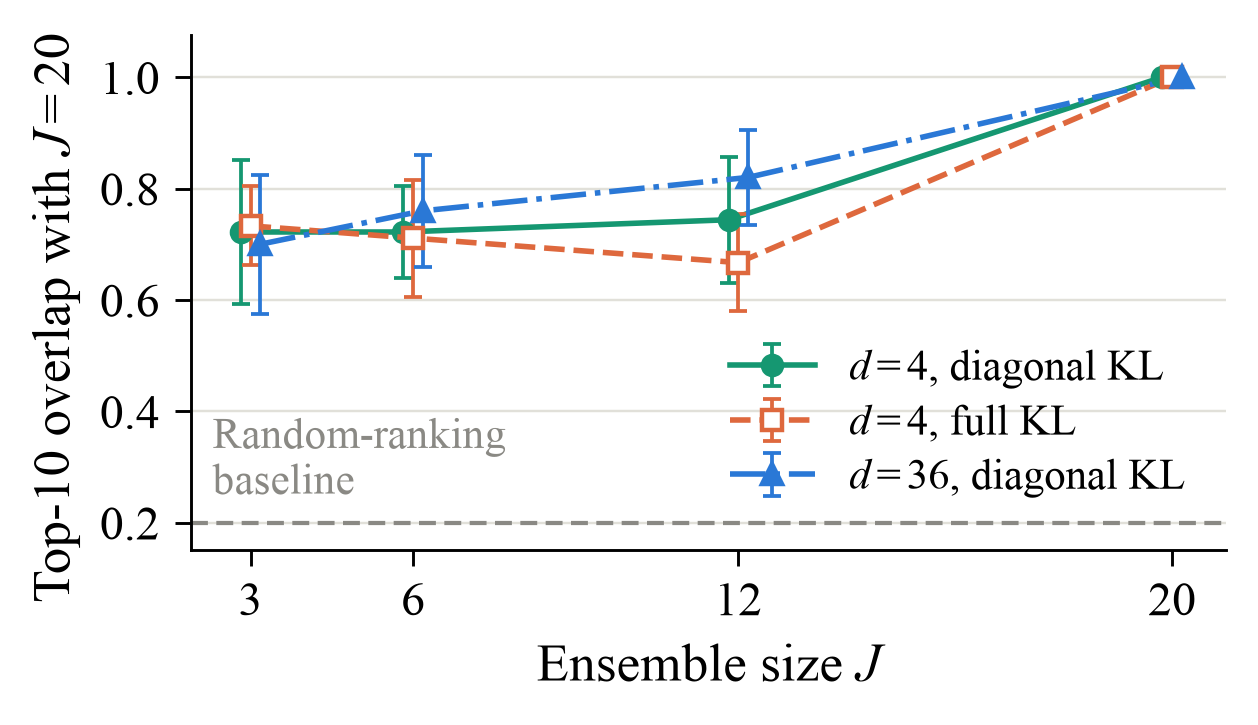}
        \caption{\textbf{Stable task ranking with smaller ensembles.} The top-10 overlap with the selection from the largest ensemble ($J=20$) stays mostly above 70\%. Details in Appendix~\ref{apd:diag_vs_full}.}
\label{fig:overlap_small_ens}
    \end{minipage}
\end{figure}

We test larger MAS with more parameters, while keeping the ensemble size much smaller than the parameter dimension ($6<36$, $10<100$, $15<400$). Active learning gives a consistent accuracy improvement over random training (Fig.~\ref{fig:accuracy_small_ens}). In the tested range, the ensemble size can stay $O(N)$ when the parameter dimension is $O(N^2)$, so the number of MAS rollouts is also $O(N)$. This shows \textbf{our active learning strategy extends to larger systems, remaining effective with high efficiency}.

We further examine the stability of the selected tasks. For a fixed-size system, the tasks selected with smaller ensembles largely overlap with those selected with a large ensemble, mostly by more than $70\%$ (Fig.~\ref{fig:overlap_small_ens}). A small ensemble leads to a rank-deficient covariance matrix, which can make the KL computation numerically unstable. We therefore use a diagonal KL. In a constructed full-rank case (reduce $d$ to 4), the diagonal KL gives nearly the same task selection as the full KL. Hence, \textbf{the task selection is robust to both the ensemble size and the diagonal approximation}.

\section{Conclusion, limitations, and future work}\label{sec: conclusion}

We first show that communication-structure optimization is sensitive to the choice of training tasks. We then show that active task selection improves downstream performance and stability, achieves better overall cost efficiency, and generalizes across diverse benchmarks. We develop an efficient and derivative-free active learning framework for this purpose, estimating task information gain with EKI. It is particularly suitable for black-box MAS in which direct information gain estimation is difficult, and it remains effective with ensembles much smaller than the parameter dimension. These results support treating task selection as an explicit optimization problem, rather than relying on randomly sampled training tasks.

\textbf{Limitations.} First, the EKI-based KL is a low-cost proxy for the information gain and is exact only under linear-Gaussian assumptions. In this non-linear problem, we assess it through task selection and downstream performance, not against the true information gain. In future work, we will try to bound its estimation error. Second, we show that active selection is more cost-efficient overall, but we do not identify its break-even point. That is, we do not answer when active selection should be preferred over random training. Third, we work on task selection from a pool of task–answer pairs. Extending it to settings that require label acquisition or task generation is left for future work.

\textbf{Summary of contents in the appendix.} First, we show the connection between EKI and score-based posterior sampling (see Appendix~\ref{apd:eki_score_da_connection}). Second, method-related details are provided, including the algorithm (\ref{apd: algorithm}), notation summary (\ref{apd:notation}), prediction-step reuse for batch utility evaluation (\ref{apd:batch_reuse}), and multi-objective utility estimation (\ref{apd: multi obj}). Then, the experimental setup (\ref{apd: details}) and MAS details (\ref{apd:mas_setup}) are provided. Additional results are reported, including scalability (\ref{apd:scale_backbone}), task examples and EKI diagnostics (\ref{apd: additional results}), statistical analyses (\ref{apd:stat_summary}), sensitivity studies (\ref{apd:sensitivity}), other active learning methods (\ref{apd: other methods}), and ablation studies (\ref{apd:ablation}). Finally, Appendix~\ref{apd:scope_generalizability} discusses failure modes and mitigations, scope, and generalizability. 

\subsection*{AI use statement}

We primarily used generative AI tools to improve the readability and polish the language of the manuscript. We also used these tools to assist with the implementation of baseline methods in our experiments and the qualitative characterization of illustrative task examples reported in Table \ref{tab:eki_task_qualitative}.

\subsection*{Reproducibility statement}

Experimental details and parameter settings are provided in Appendix~\ref{apd: details} and \ref{apd:mas_setup} to facilitate reproduction of our results. The code used in this work will be released upon acceptance.

\bibliography{reference}
\bibliographystyle{iclr2027_conference}
\appendix
\include{appendix}

\end{document}

%% file: appendix.tex
\section{Methodology details and extensions}

\subsection{EKI and score-based posterior sampling}
\label{apd:eki_score_da_connection}

In this appendix, we share some insights into the connection between the EKI update used in this work and score-based posterior sampling. We also discuss the limitations of EKI, which can guide future studies.

Note that, in the main text, we write the Bayesian update abstractly as \(p(\mathbf z\mid q,y)\), without specifying how the compatibility between the MAS output $g$ and the reference answer \(y\) is represented. In the EKI implementation, this relationship is made explicit through the performance score \(l=\phi(g,q,y)\) and the desired performance value  \(l^*\). We therefore express the posterior update below as $p(\mathbf{z}|l^*)$ in the generic derivation, or as $p(\mathbf{z}|l^*,q)$ when the dependence on a specific task $q$ is made explicit.

\subsubsection{Score-based posterior sampling}

The score-based posterior sampling targets the Bayesian update:
\begin{equation}
    p(\mathbf{z}|l^*) \propto p(\mathbf{z})p(l^*|\mathbf{z}),
\end{equation}
which is handled in the form of score functions to avoid explicitly accounting for the computationally expensive normalization constant in the above Bayesian update:
\begin{equation}
    \nabla_\mathbf{z} \log p(\mathbf{z}|l^*) = \nabla_\mathbf{z} \log p(\mathbf{z}) +\nabla_\mathbf{z} \log p(l^*|\mathbf{z}).
    \label{eq: apd posterior score}
\end{equation}
By working with the score function of the posterior distribution, sampling can be achieved through either classical Langevin dynamics~\cite{roberts1996exponential,song2019generative} or more recent score-based diffusion models~\cite{song2020score,chung2022diffusion}. It is worth noting that diffusion models rely on a series of score functions that correspond to the original one with different levels of noise injected, which has been demonstrated to be more robust than merely relying on the targeted score function.

\subsubsection{Connection between score-based posterior sampling and EKI}

\paragraph{Observation model.} For a task \(q\), let $\mathcal{H}_q(\mathbf{z})$ denote the forward model associated with the scalar performance score \(l\). In the notation of the main text, \(\mathcal{H}_q(\mathbf{z})\) is a combination of $\phi$ and $G$. Consider the Gaussian observation model
\begin{equation}
    l = \mathcal{H}_q(\mathbf{z})+\eta,
    \qquad
    \eta\sim\mathcal{N}(0,\Gamma),
    \label{eq:apd_obs_model}
\end{equation}
where \(\Gamma\) is the observation noise covariance in score space. Following Eq.~\eqref{eq:apd_obs_model}, given a desired value of performance score $l^*$ (the observation), the likelihood is
\begin{equation}
    p(l^*|\mathbf{z},q)
    \propto
    \exp\left(
    -\frac{1}{2}
    \bigl(l^*-\mathcal{H}_q(\mathbf{z})\bigr)^\top
    \Gamma^{-1}
    \bigl(l^*-\mathcal{H}_q(\mathbf{z})\bigr)
    \right).
    \label{eq:apd_likelihood}
\end{equation}

\paragraph{Likelihood-score.} If \(\mathcal{H}_q\) is differentiable, then
\begin{equation}
    \nabla_{\mathbf{z}}\log p(l^*|\mathbf{z},q)
    =
    J_q(\mathbf{z})^\top
    \Gamma^{-1}
    \bigl(l^*-\mathcal{H}_q(\mathbf{z})\bigr),
    \label{eq:apd_likelihood_score}
\end{equation}
where
\begin{equation}
    J_q(\mathbf{z})=\nabla_{\mathbf{z}}\mathcal{H}_q(\mathbf{z})
\end{equation}
is the Jacobian of the forward model. Combining the likelihood score with the prior score gives
\begin{equation}
    \nabla_{\mathbf{z}}\log p(\mathbf{z}|q,l^*)
    =
    \nabla_{\mathbf{z}}\log p(\mathbf{z})
    +
    J_q(\mathbf{z})^\top
    \Gamma^{-1}
    \bigl(l^*-\mathcal{H}_q(\mathbf{z})\bigr).
    \label{eq:apd_posterior_score}
\end{equation}


\paragraph{Local linear-Gaussian approximation induced by the ensemble.} Assume the initial ensemble is a local Gaussian prior,
\begin{equation}
    \mathbf{z}\sim \mathcal{N}(\bar{\mathbf{z}},C^{zz}), \quad  \bar{\mathbf{z}}
    =
    \frac{1}{J}\sum_{j=1}^{J}\mathbf{z}^{(j)}.
\end{equation}
Around the initial ensemble mean \(\bar{\mathbf{z}}\), we approximate the forward model by
\begin{equation}
    \mathcal{H}_q(\mathbf{z})
    \approx
    \mathcal{H}_q(\bar{\mathbf{z}})
    +
    G_q(\mathbf{z}-\bar{\mathbf{z}}),
    \qquad
    G_q=J_q(\bar{\mathbf{z}}).
    \label{eq:apd_local_linear}
\end{equation}
Under this local linear-Gaussian model, we define a local posterior covariance as
\begin{equation}
    C_{\mathrm{loc}}
    =
    \left[
    (C^{zz})^{-1}
    +
    G_q^\top\Gamma^{-1}G_q
    \right]^{-1},
    \label{eq:apd_Cloc}
\end{equation}
in order to write the Kalman gain matrix in an explicit form involving $G_q^\top \Gamma^{-1}$
\begin{equation}
    K_q
    = C^{zl}(C^{ll}+\Gamma)^{-1} \approx
    C^{zz}G_q^\top
    \bigl(G_qC^{zz}G_q^\top+\Gamma\bigr)^{-1}=C_{\mathrm{loc}}G_q^\top \Gamma^{-1}.
    \label{eq:apd_Kalman_gain}
\end{equation}
The detailed derivation of Eq.~\eqref{eq:apd_Kalman_gain} is summarized in Appendix~\ref{apd: proof}. Then the EKI update used in the main text is
\begin{equation}
    \mathbf{z}^{(j)}_{\mathrm{post}}
    =
    \mathbf{z}^{(j)}
    +
  C_{\mathrm{loc}}G_q^\top \Gamma^{-1}
    \bigl(l^*-\mathcal{H}_q(\mathbf{z}^{(j)})\bigr),
    \label{eq:apd_eki_update}
\end{equation}
The second term on the right-hand side is similar to the second term in Eq.~\eqref{eq:apd_posterior_score}, which can be shown more clearly as below:
\begin{equation}
\begin{aligned}
    C_{\mathrm{loc}}G_q^\top\Gamma^{-1}
    \bigl(l^*-\mathcal{H}_q(\mathbf{z}^{(j)})\bigr) 
    &\approx C_{\mathrm{loc}} \left[ J_q(\mathbf{z})^\top
    \Gamma^{-1}
    \bigl(l^*-\mathcal{H}_q(\mathbf{z})\bigr) \right] \\
    &=
    C_{\mathrm{loc}}
    \left[
    \nabla_{\mathbf{z}}\log p(l^*|\mathbf{z},q)
    \right]_{\mathbf{z}=\mathbf{z}^{(j)}}.
\label{eq:apd_eki_as_score_step}
\end{aligned}
\end{equation}
The equality is obtained if the forward model is linear, i.e., $J_q=G_q$. Equation~\eqref{eq:apd_eki_as_score_step} reveals a connection between EKI and score-based posterior transport -- the update formula in EKI is equivalent to applying the local posterior covariance \(C_{\mathrm{loc}}\) as a preconditioner to the likelihood score at each ensemble particle. 

 
To summarize, EKI is computationally efficient but naturally limited by the covariance structure represented by the ensemble, whereas score-based posterior sampling is more expressive for non-Gaussian or multimodal posteriors at the cost of repeated score evaluations during sampling.

\subsubsection{Detailed derivation of Eq.~\eqref{eq:apd_Kalman_gain}}\label{apd: proof}

\textbf{1. Justification of the first approximation} 

Consider
\begin{equation}
    \begin{aligned}
        C^{zl}&=\frac{1}{J-1}\sum_{j=1}^J (\mathbf{z}^{(j)}-\bar{\mathbf{z}})(l^{(j)}-\bar{l})^\top\\
        &\approx \frac{1}{J-1}\sum_{j=1}^J (\mathbf{z}^{(j)}-\bar{\mathbf{z}})\left[G_q (\mathbf{z}^{(j)}-\bar{\mathbf{z}}) \right]^\top\\
        &=\frac{1}{J-1}\sum_{j=1}^J (\mathbf{z}^{(j)}-\bar{\mathbf{z}}) (\mathbf{z}^{(j)}-\bar{\mathbf{z}})^\top G_q^\top\\
        &=C^{zz} G_q^\top.
    \end{aligned}
\end{equation}

Similarly, it can be shown that $C^{ll}\approx G_q C^{zz} G_q^\top$. Therefore, the first approximation holds, and the approximation will be exact when the forward model is linear.

\textbf{2. Proof of the final equality}

Here we show
\begin{equation}
    C^{zz}G_q^\top
    \bigl(G_qC^{zz}G_q^\top+\Gamma\bigr)^{-1}  =C_{\mathrm{loc}}G_q^\top \Gamma^{-1}.
\end{equation}

We first define $S=G_qC^{zz}G_q^\top+\Gamma$ for notational simplicity. Starting from the right-hand side,
\begin{equation}
    \begin{aligned}
        C_{\mathrm{loc}}G_q^\top \Gamma^{-1}
        & = C_{\mathrm{loc}}G_q^\top \Gamma^{-1}SS^{-1}\\
        &=C_{\mathrm{loc}}G_q^\top \Gamma^{-1}(G_qC^{zz}G_q^\top+\Gamma)S^{-1}\\
        &=C_{\mathrm{loc}} \left[ G_q^\top \Gamma^{-1}G_qC^{zz}G_q^\top S^{-1}+ G_q^\top \Gamma^{-1}\Gamma S^{-1}\right]\\
        &=C_{\mathrm{loc}} \left[ G_q^\top \Gamma^{-1}G_qC^{zz}G_q^\top S^{-1}+ G_q^\top S^{-1}\right]\\
        &=C_{\mathrm{loc}} \left[ G_q^\top \Gamma^{-1}G_qC^{zz}G_q^\top S^{-1}+ (C^{zz})^{-1}C^{zz}G_q^\top S^{-1}\right]\\
        &=C_{\mathrm{loc}}  \left[ G_q^\top \Gamma^{-1}G_q+ (C^{zz})^{-1} \right]C^{zz}G_q^\top S^{-1}\\
        &=C_{\mathrm{loc}} C_{\mathrm{loc}}^{-1} C^{zz}G_q^\top S^{-1}\\
        &= C^{zz}G_q^\top S^{-1}.\\
    \end{aligned}
\end{equation}

\subsubsection{Limitations of EKI}
In this work, we use EKI as an ensemble-based method to efficiently approximate the Bayesian update involved in the evaluation of information gain. It is much more efficient than diffusion models and other variants of stochastic interpolants, such as flow matching, since it avoids solving a multi-step reverse SDE/ODE, each step of which requires a forward model evaluation. The number of inference steps for diffusion models is generally much higher than that of EKI. 

As a trade-off, EKI cannot generate samples from arbitrarily complex posteriors. Theoretically, EKI is exactly equivalent to score-based posterior sampling only under linear-Gaussian assumptions. Although many works have shown EKI can handle non-linear problems in practice, one should not expect a perfect match between the EKI-approximated posterior and the true one that is highly non-Gaussian. The posterior generated by EKI can be viewed as an optimal Gaussian approximation to the true posterior. In this work, we have shown that such an approximate posterior can be useful for active learning and, more specifically, Bayesian experimental design problems.

\subsection{Algorithm}\label{apd: algorithm}

We provide the overall algorithm of active learning in Algorithm~\ref{alg:active_selection} and also the detailed algorithm to evaluate the EKI-based utility in Algorithm~\ref{alg:eki_utility}. For multi-step EKI, one can repeat Lines 2--10 using the updated ensemble from the previous step as the current ensemble, and then compute the information gain from the final updated ensemble.

\begin{algorithm}[H]
\caption{Active Task Selection for Communication-Graph Optimization}
\label{alg:active_selection}
\begin{algorithmic}[1]
\Require Training pool $\mathcal{D}_{\mathrm{tr}}=\{(q_i,y_i)\}_{i=1}^{M}$; representative-pool size $R$; utility-evaluation budget $K_{\mathrm{eval}}$; final training budget $K$; ensemble size $J$; prior distribution $p(\mathbf{z})$; target score $l^*$; observation-noise covariance $\Gamma$.
\Ensure Selected task subset $\mathcal{S}$.

\State Embed each task $q_i$ into a vector $v_i = v(q_i)$.
\State Construct a representative pool $\mathcal{P}_R \subseteq \mathcal{D}_{\mathrm{tr}}$ with $|\mathcal{P}_R|=R$ using greedy farthest-point selection in the embedding space.
\State Initialize the evaluated set $\mathcal{O}\gets \emptyset$.
\State Select an initial subset $\mathcal{B}_0 \subseteq \mathcal{P}_R$ for utility evaluation.

\For{each task $(q,y)\in \mathcal{B}_0$}
    \State Estimate its utility $U(q)$ by Algorithm~\ref{alg:eki_utility}.
    \State Update $\mathcal{O}\gets \mathcal{O}\cup\{(q,U(q))\}$.
\EndFor

\While{the total number of evaluated tasks is smaller than $K_{\mathrm{eval}}$}
\State Fit or refit the surrogate model $\bar{\psi}: v(q)\mapsto U(q)$ on the evaluated set $\mathcal{O}$, where $\bar{\psi}$ is implemented by PLS dimension reduction followed by GP regression.
    \State Use the surrogate acquisition rule, e.g., Thompson sampling, to select a new batch
    \[
        \mathcal{B}\subseteq \mathcal{P}_R\setminus \{q:(q,U(q))\in\mathcal{O}\}.
    \]
    \For{each task $(q,y)\in \mathcal{B}$}
        \State Estimate its utility $U(q)$ by Algorithm~\ref{alg:eki_utility}.
        \State Update $\mathcal{O}\gets \mathcal{O}\cup\{(q,U(q))\}$.
    \EndFor
\EndWhile

\State Select the final training subset
\[
    \mathcal{S}
    =
    \operatorname{TopK}_{(q,U(q))\in\mathcal{O}} U(q).
\]
\State \Return $\mathcal{S}$.
\end{algorithmic}
\end{algorithm}

\begin{algorithm}[t]
\caption{One-Step EKI Utility Estimation for a Task}
\label{alg:eki_utility}
\begin{algorithmic}[1]
\Require Task-answer pair $(q,y)$; prior distribution $p(\mathbf{z})$; ensemble size $J$; target score $l^*$; observation-noise covariance $\Gamma$.
\Ensure Task utility $U(q)$.

\State Draw an initial ensemble
\[
    \{\mathbf{z}^{(j)}\}_{j=1}^{J}\sim p(\mathbf{z}).
\]

\For{$j=1,\dots,J$}
    \State Run the MAS forward map:
    \[
        g^{(j)} = G(q;\mathbf{z}^{(j)},\omega^{(j)}).
    \]
    \State Evaluate the MAS output:
    \[
        l^{(j)}=\phi(g^{(j)},q,y).
    \]
\EndFor

\State Compute empirical means
\[
    \bar{\mathbf{z}}=\frac{1}{J}\sum_{j=1}^{J}\mathbf{z}^{(j)},
    \qquad
    \bar{l}=\frac{1}{J}\sum_{j=1}^{J}l^{(j)}.
\]

\State Compute the empirical cross-covariance and observation (score) covariance:
\[
    C^{zl}
    =
    \frac{1}{J-1}
    \sum_{j=1}^{J}
    \bigl(\mathbf{z}^{(j)}-\bar{\mathbf{z}}\bigr)
    \bigl(l^{(j)}-\bar{l}\bigr)^\top,
\]
\[
    C^{ll}
    =
    \frac{1}{J-1}
    \sum_{j=1}^{J}
    \bigl(l^{(j)}-\bar{l}\bigr)
    \bigl(l^{(j)}-\bar{l}\bigr)^\top.
\]

\For{$j=1,\dots,J$}
    \State Update the ensemble particle:
    \[
        \mathbf{z}^{(j)}_{\mathrm{post}}
        =
        \mathbf{z}^{(j)}
        +
        C^{zl}
        \bigl(C^{ll}+\Gamma\bigr)^{-1}
        \bigl(l^*-l^{(j)}\bigr).
    \]
\EndFor

\State Approximate the information gain using a Gaussian approximation of the ensembles:
\[
    U(q)
    =
    I(q)
    \approx
    D_{\mathrm{KL}}
    \left(
    \{\mathbf{z}^{(j)}_{\mathrm{post}}\}_{j=1}^{J}
    \,\middle\|\,
    \{\mathbf{z}^{(j)}\}_{j=1}^{J}
    \right).
\]

\State \Return $U(q)$.
\end{algorithmic}
\end{algorithm}

\subsection{Notation summary}\label{apd:notation}

We provide a notation summary in Table~\ref{tab:notation}.

\begin{table}[t]
\caption{Summary of key notation.}
\label{tab:notation}
\centering
\begin{small}
\begin{tabular}{p{0.22\linewidth}p{0.70\linewidth}}
\toprule
Symbol & Meaning \\
\midrule

$\mathcal{D}_{\mathrm{tr}}=\{(q_i,y_i)\}_{i=1}^{M}$ 
& Full training task pool, where $q_i$ is a task and $y_i$ is its ground-truth answer. \\

$\mathcal{S}$ 
& Selected subset of tasks used for communication-graph optimization. \\


$\mathcal{G}=(V,E^S,E^T)$
& Multi-agent communication graph, where $V$ is the set of agents and $E^S,E^T$ are the spatial and temporal adjacency matrices. \\

$\mathbf{z}=\{Z^S,Z^T\}$ 
& Trainable graph logits that parameterize the spatial and temporal communication masks. \\

$A=\{A^S,A^T\}$ 
& Random communication mask sampled from the logits $\mathbf{z}$. \\

$G(q;\mathbf{z},\omega)$ 
& MAS forward map for task $q$ under graph logits $\mathbf{z}$ and internal stochasticity $\omega$. \\

$g$ 
& Output of the MAS, such as a choice, numerical answer, or code output. \\

$\phi(g,q,y)$ 
& Evaluator that maps the MAS output, task, and ground truth to a performance score. \\

$l$ 
& Performance score produced by the evaluator. \\

$U(q)$ 
& Utility of a task $q$, measured by the information gain induced by the task. \\

$U(\mathcal{S})$ 
& Utility of a selected task subset $\mathcal{S}$. \\

$I(q)$ 
& Information gain of a single task, approximated by the KL divergence between the EKI-updated ensemble and the prior ensemble. \\

$p(\mathbf{z})$ 
& Prior distribution over graph logits. \\

$p(\mathbf{z}\mid q,y)$ 
& Posterior distribution over graph logits after conditioning on a task-answer pair $(q,y)$. \\

$D_{\mathrm{KL}}(\cdot\|\cdot)$ 
& Kullback--Leibler divergence used to quantify information gain. \\

$\{\mathbf{z}^{(j)}\}_{j=1}^{J}$ 
& EKI ensemble representing uncertainty over graph logits. \\

$J$ 
& EKI ensemble size. \\

$l^*$ 
& Desired target score used in the EKI update. \\

$\Gamma$ 
& Observation-noise covariance (in score space) in the EKI update.\\

$v(q)$ 
& Embedding vector of task $q$. \\

$\mathcal{P}_R$ 
& Representative task pool selected from the original training pool. \\

$\bar{\psi}$ 
& Surrogate model, e.g., a GP, that approximates the map from reduced task embeddings to utilities. \\

\bottomrule
\end{tabular}
\end{small}
\end{table}

\subsection{Prediction-step reuse under exhaustive subset evaluation}\label{apd:batch_reuse}

This appendix presents a possible batch extension of the single-task EKI-based utility estimator. This formulation is not used in the main experiments, where task gains are computed at the single-task level. We include it to show how the same EKI-based update can be generalized to subset-level information gain and how forward evaluations can be reused when candidate subsets overlap. We focus here on the estimator and the associated reuse structure, rather than on the full combinatorial search problem over task subsets. Developing a practical acquisition strategy for efficiently exploring the batch-subset space is an interesting direction for future work.

\subsubsection{Reuse formulation in batch EKI}

We describe how the prediction-step reuse is incorporated into the batch EKI formulation. Consider a candidate pool $\mathcal{D}_\text{tr}=\{(q_i,y_i)\}_{i=1}^M$ and the initial ensemble $\{\mathbf{z}^{(j)}\}_{j=1}^J$. For each task $q_i$ and each ensemble particle $\mathbf{z}^{(j)}$, we first run the forward map and compute the corresponding score
\[
    g_i^{(j)} = G(q_i;\mathbf{z}^{(j)}),
    \qquad
    l_i^{(j)} = \phi(g_i^{(j)},q_i,y_i) \in \mathbb{R}.
\]
These task-level scores are cached once and reused for different candidate subsets. Specifically, for each ensemble particle, we define the cached score vector over the full candidate pool as
\[
    \mathbf{l}^{(j)}
    =
    [l_1^{(j)},l_2^{(j)},\dots,l_M^{(j)}]^\top
    \in \mathbb{R}^M.
\]

For a candidate subset $\mathcal{S}=\{q_{i_1},q_{i_2},\dots,q_{i_m}\}$, the batch score vector for particle $j$ is obtained by selecting the corresponding entries from the cached pool-level score vector:
\[
    \mathbf{l}_{\mathcal{S}}^{(j)}
    =
    [l_{i_1}^{(j)},l_{i_2}^{(j)},\dots,l_{i_m}^{(j)}]^\top
    \in \mathbb{R}^m.
\]
Thus, forming $\mathbf{l}_{\mathcal{S}}^{(j)}$ only requires indexing and stacking cached task-level scores, rather than rerunning the forward map.

The subset-specific ensemble mean of the score vectors is
\[
    \bar{\mathbf{l}}_{\mathcal{S}}
    =
    \frac{1}{J}\sum_{j=1}^J \mathbf{l}_{\mathcal{S}}^{(j)},
    \qquad
    \bar{\mathbf{z}}
    =
    \frac{1}{J}\sum_{j=1}^J \mathbf{z}^{(j)}.
\]
Then the cross-covariance and score covariance used in the EKI update are
computed as
\[
    C_{\mathbf{z}l}^{\mathcal{S}}
    =
    \frac{1}{J-1}
    \sum_{j=1}^J
    (\mathbf{z}^{(j)}-\bar{\mathbf{z}})
    (\mathbf{l}_{\mathcal{S}}^{(j)}-\bar{\mathbf{l}}_{\mathcal{S}})^\top,
\]
and
\[
    C_{ll}^{\mathcal{S}}
    =
    \frac{1}{J-1}
    \sum_{j=1}^J
    (\mathbf{l}_{\mathcal{S}}^{(j)}-\bar{\mathbf{l}}_{\mathcal{S}})
    (\mathbf{l}_{\mathcal{S}}^{(j)}-\bar{\mathbf{l}}_{\mathcal{S}})^\top .
\]
Here, $C_{\mathbf{z}l}^{\mathcal{S}}\in\mathbb{R}^{d_z\times m}$ and $C_{ll}^{\mathcal{S}}\in\mathbb{R}^{m\times m}$, where $d_z$ is the dimension of the graph parameter vector.

Given the batch-level target score vector $\mathbf{l}_{\mathcal{S}}^{\mathrm{obs}}\in\mathbb{R}^m$ and the observation noise covariance $\Gamma_{\mathcal{S}}\in\mathbb{R}^{m\times m}$, the EKI update for subset $\mathcal{S}$ is
\[
    \mathbf{z}_{\mathrm{post},\mathcal{S}}^{(j)}
    =
    \mathbf{z}^{(j)}
    +
    C_{\mathbf{z}l}^{\mathcal{S}}
    \left(
        C_{ll}^{\mathcal{S}}+\Gamma_{\mathcal{S}}
    \right)^{-1}
    \left(
        \mathbf{l}_{\mathcal{S}}^{\mathrm{obs}}
        -
        \mathbf{l}_{\mathcal{S}}^{(j)}
    \right).
\]
The resulting posterior ensemble $\{\mathbf{z}_{\mathrm{post},\mathcal{S}}^{(j)}\}_{j=1}^J$ is then used to compute the batch information gain $I(\mathcal{S})$.

This formulation shows that the prediction step is performed at the task level, while the EKI update step is performed at the subset level. Therefore, when different candidate subsets overlap, their shared tasks reuse the same cached scores $l_i^{(j)}$, whereas the covariance matrices
$C_{\mathbf{z}l}^{\mathcal{S}}$, $C_{ll}^{\mathcal{S}}$, the posterior ensemble, and the resulting information gain remain specific to each subset $\mathcal{S}$.

\subsubsection{Cost analysis}

If batch information-gain evaluation involves exhaustive subset enumeration, the reuse strategy can substantially reduce repeated forward-model evaluations. Consider a candidate pool of $M$ tasks and suppose that all size-$m$ subsets are evaluated under the same current ensemble of size $J$. A naive exhaustive evaluation would compute the prediction step separately for each subset, requiring
\[
    \binom{M}{m}mJ
\]
forward-model evaluations. Here, $\binom{M}{m}$ denotes the number of possible size-$m$ subsets that can be formed from a pool of $M$ candidate tasks. However, the prediction associated with a task does not depend on the subset in which the task appears, as long as all subsets are evaluated from the same current ensemble. Therefore, task-level predictions can be cached once for all $M$ unique tasks, reducing the prediction-step cost to
\[
    MJ.
\]
The resulting reduction factor is
\[
    \frac{\binom{M}{m}mJ}{MJ}
    =
    \binom{M-1}{m-1}.
\]

For example, selecting $5$ tasks from a pool of $10$ already gives $\binom{10}{5}=252$ candidate subsets, resulting in a $\binom{9}{4}=126$-fold reduction in prediction-step forward evaluations.
Selecting $10$ tasks from a pool of $50$ gives
\[
    \binom{50}{10}=10{,}272{,}278{,}170,
\]
which is on the order of $10^{10}$ possible subsets. The corresponding prediction-step reduction factor is
\[
    \binom{49}{9}=2{,}054{,}455{,}634.
\]

This large factor comes from the combinatorial repetition of the same tasks across different subsets. These values are meant only to illustrate the potential savings under exhaustive subset enumeration, since our method does not enumerate all possible subsets in practice. Nevertheless, the same reuse
principle applies whenever the evaluated candidate subsets overlap: shared tasks can reuse cached prediction results, reducing the number of repeated forward rollouts.

\subsubsection{Remark}

One subtlety of aggressive prediction reuse is that all candidate subsets are evaluated conditional on the same finite ensemble. In the large-ensemble limit, this is harmless, since the empirical ensemble distribution converges to the target uncertainty distribution and reuse only avoids duplicated forward evaluations. With a small ensemble, however, the ensemble provides only a
low-rank and random representation of parameter uncertainty. Consequently, all subset utilities are evaluated through the same empirical covariance directions and share the same ensemble-induced approximation error. This might make the ranking of candidate subsets sensitive to the particular ensemble realization: subsets that are informative along the sampled directions may be favored, while
subsets whose informativeness lies in poorly represented directions may be undervalued. The severity of this effect depends on the ensemble size, the posterior geometry, and the diversity of the candidate subsets. A systematic characterization of this effect is left for future work.

However, since reuse reduces repeated forward rollouts, part of the saved computational budget could be allocated to increasing the ensemble size. Besides, in practice, one may reuse task-level predictions only within a limited collection of candidate subsets. Designing finite-reuse strategies that balance computational savings and finite-ensemble approximation error is also an interesting direction for future work.

Overall, this formulation highlights a key advantage of the batch EKI view: the expensive prediction step can be organized at the task level, while subset-specific information gains are obtained through lightweight recombination in the update step.

\subsection{Multi-objective extension of EKI-based utility estimation}\label{apd: multi obj}

\subsubsection{Vector-valued utility estimation}

It is also possible to incorporate multiple objectives explicitly, rather than collapsing them into a single scalar score. In a traditional scalarized multi-objective formulation, the loss is often written as
\begin{equation}
    l = \phi(\cdot) = \alpha_1 l_1 + \alpha_2 l_2 + \cdots + \alpha_n l_n \in \mathbb{R},
\end{equation}
where $l_i$ denotes the $i$-th objective and $\alpha_i$ is its corresponding weight. In graph optimization, these objectives may represent, for example, downstream accuracy, graph sparsity, and regularization~\cite{zhang2024cut,wang2025agentdropout}. However, such scalarization mixes the objective signals before the EKI update is formed, so improvement in one component may offset degradation in another.

Instead, we can define a vector-valued pseudo-observation as
\begin{equation}
    l= [\alpha_1 l_1, \alpha_2 l_2, \cdots, \alpha_n l_n]^\top \in \mathbb{R}^n,
\end{equation}
so that each objective enters the update as a separate component. Under the standard EKI formulation, this corresponds to matching a weighted quadratic residual in the observation space, which allows the posterior update to respond to each objective more explicitly instead of only their pre-aggregated sum. As a result, the update is driven by the component-wise residual structure rather than by a single pre-aggregated scalar score. This does not remove trade-offs among objectives, but it avoids premature cancellation caused by scalarization and yields a more structured update signal.

\subsubsection{Remark}

The multi-objective extension (Appendix~\ref{apd: multi obj}) and the batch extension (in Sec~\ref{sec: batch}) act on different axes of the pseudo-observation and are therefore complementary rather than redundant. The batch extension increases the observation dimension across tasks, while the multi-objective extension increases the dimension within each task by keeping different objectives as separate components. When both are used together, each task $q_i$ is associated with an $n$-dimensional pseudo-observation vector
\begin{equation}
    l_i = [\alpha_1 l_{i,1}, \alpha_2 l_{i,2}, \cdots, \alpha_n l_{i,n}]^\top \in \mathbb{R}^n,
\end{equation}
and a batch $\mathcal{S}=\{q_i,y_i\}_{i=1}^m$ is represented by concatenating these task-wise vectors into a single observation vector
\begin{equation}
    l_{\mathcal{S}} = [l_1^\top, l_2^\top, \cdots, l_m^\top]^\top \in \mathbb{R}^{mn}.
\end{equation}
This unified form can be directly used in the EKI update, allowing the method to simultaneously account for multiple tasks and multiple objectives.


\section{Experimental details}\label{apd: details}

In this section, we provide detailed experimental settings for the main MMLU and GSM8K experiments, followed by the settings for the broader benchmarks in Appendix~\ref{apd:broader_benchmark_settings}.

\subsection{Dataset}

For GSM8K, we use the \texttt{train} split as the task pool for communication-structure optimization and the \texttt{test} split for evaluation. We report results on the first 460 tasks in the \texttt{test} split, which provides a sufficiently large evaluation set for stable empirical comparison across runs.

For MMLU, the original implementation in~\cite{zhang2024cut} uses the `dev` split to randomly select tasks for graph optimization. However, the `dev` split is less suitable as an active-learning candidate pool, since it contains only five tasks for each subject. Such a small and highly structured pool limits task diversity and is less favorable for representative selection based on clustering or task embeddings. Therefore, in our setting, we use the `test` split as the candidate pool for communication-structure optimization, since it is large enough to support active learning over a more diverse set of tasks. For evaluation, we use the first 153 tasks in the \texttt{val} split, following~\cite{zhang2024cut}. In this way, the task pool used for communication-structure optimization is separated from the evaluation set.

Unless otherwise stated, all the other settings are the same for both datasets.

\subsection{Active learning pipeline and random training pipeline}

For active learning, we use the same selection pipeline in both datasets and settings (with and without attack). We first randomly sample 1000 tasks from the full training set, then select 50 representative tasks based on task embeddings, and finally choose training tasks from this 50-task pool.

The 1000-to-50-to-10 pipeline is used as an illustrative default configuration, rather than as a tuned or problem-specific choice. The final number (10) of selected tasks is determined by the training budget (i.e., how many task usages one can afford during graph optimization). The size of the intermediate candidate pool is determined by the selection budget (i.e., how many tasks can be evaluated for informativeness). In practice, these three quantities can be adjusted independently to accommodate different computational settings and resource constraints.

For active learning, we select 10 informative tasks and use each task twice during training, resulting in 20 total task usages. Here, a “task usage” means one pass where the LLM-MAS is executed on a task to produce an answer, the answer is scored by the task-specific metric, and this score is used to compute the training signal (e.g., a policy-gradient estimate) for updating the communication-graph parameters. We adopt this repeated-use design because using each selected task only once may not provide a sufficiently reliable training signal. Due to stochasticity in the forward map and the resulting noisy graph-optimization signal, the effect of an informative task can be weakened if it is used only once.

For random selection, we randomly select tasks from the whole training pool and match the total number of task usages in training. On GSM8K, for consistency with the active-learning training schedule, we randomly sample 10 tasks and use each task twice. On MMLU, we notice that the original training implementation of~\cite{zhang2024cut} uses each selected task only once. To keep the forward model and training procedure as black-boxes with minimal changes, we follow this design and randomly sample 20 tasks from the whole pool, using each once, so that the total number of task usages remains 20. Overall, across the two commonly used ways to allocate an equal training budget, we consistently observe that active learning outperforms random task training.

In addition, on MMLU we performed a sanity check by swapping the random-training repetition schedule from 20 tasks × 1 to 10 tasks × 2 while keeping the total task usages fixed. The random baseline remains very similar ($76.87 \pm 3.72$ as shown in the main text vs. $76.42 \pm 2.55$ over 15 runs), indicating that this choice does not materially affect the random-training performance.

\subsection{EKI hyperparameters}

We use a one-step EKI update with ensemble size 6, prior perturbation standard deviation 2.0, observation noise 0.1, and damping factor 0.7.

Since our experiments use a single interaction round, temporal communication is inactive: \(E^T\) is fixed to the zero matrix and is not optimized. Therefore, only the spatial graph logits \(Z^S\in\mathbb{R}^{6\times6}\) are optimized, resulting in a 36-dimensional graph-parameter space. An ensemble size of 6 is small relative to the parameter dimension. We use a prior perturbation standard deviation of 2.0 to cover a relatively broad range of initial perturbations. Since the forward system does not include explicit observation noise, we use a small fixed value of 0.1 as a regularizing noise level in the EKI update. The damping factor 0.7 is used to make the update more conservative and numerically stable.

For the performance score $l$ and target value $l^*$, we use the log-probability of the correct answer (A/B/C/D) for MMLU and therefore $l^*=0$. We use 0/1 correctness for GSM8K and set $l^*=1$. 

\subsection{KL divergence estimation}

For the empirical KL computation, we use a diagonal Gaussian approximation rather than a full-covariance Gaussian approximation. This choice is motivated by the small ensemble size used in our experiments. With $J=6$ particles and a 36-dimensional graph-logit parameter, the empirical full covariance matrix would be rank-deficient and therefore unsuitable for a standard Gaussian KL computation. We therefore estimate only the coordinate-wise empirical means and variances of the prior and updated ensembles. We also apply a small variance floor of $10^{-8}$ to avoid numerical instability from zero or near-zero empirical variances.

This diagonal approximation is an implementation choice for the small-ensemble setting, rather than a restriction of the proposed framework. When the ensemble size is large relative to the parameter dimension and the covariance estimate is well-conditioned, we recommend using a full-covariance Gaussian KL instead. This is consistent with similar ensemble-based Gaussian KL estimation practice in \cite{yang2025bayesian}.

\subsection{Thompson sampling and Gaussian process surrogate}\label{apd: ts}

After representative selection, Thompson sampling is performed on the 50-task pool in a sequential batched manner. We first randomly evaluate 10 or 12 tasks to form an initial observed set. Then we perform additional acquisition rounds, selecting 5 or 6 new tasks in each round, giving an overall schedule of $10+5\times 3$ for MMLU and $12+6\times 4$ for GSM8K. After each round, the newly evaluated tasks and their true utility values are added to the observed set, and the surrogate model is refit. The purpose of these rounds is to progressively improve the surrogate using newly acquired labeled tasks, so that later selections are guided by a more accurate model.

To make surrogate fitting more stable, we first use the 10 or 12 initially evaluated tasks and their utility values to supervise a partial least squares (PLS) projection. Concretely, the task embeddings from representative selection are originally 384-dimensional, and PLS is used to reduce them to a much smaller latent space, typically 2 or 3 dimensions. This step uses the observed utility values to identify directions in the embedding space that are most relevant to task informativeness, rather than performing unsupervised dimensionality reduction. We use a small latent dimension because the representative pool contains only 50 tasks and the number of labeled tasks is limited in the early rounds, so a higher-dimensional surrogate is not expected to be beneficial in this setting.

A Gaussian process (GP) surrogate is then fit on the reduced features to predict utility values for the remaining tasks. Thompson sampling is implemented by drawing samples from the GP posterior and selecting the tasks with the highest sampled values. In this way, PLS provides a utility-aware low-dimensional representation, while the GP supplies the predictive model used for sequential task acquisition. The GP uses a Matérn kernel with smoothness parameter $\nu=2.5$, combined with a learnable output scale and an additive white-noise term.

By applying this sequential decision-making process, we avoid evaluating the true utility of all 50 tasks in the representative pool. Instead, we only evaluate 25 tasks for MMLU and 36 tasks for GSM8K, reducing the utility-evaluation cost by 50\% and 28\%, respectively. In our experiments, this reduced evaluation budget still achieves more than 80\% overlap with the top tasks selected by full enumeration. This suggests that the surrogate-assisted Thompson-sampling procedure captures most of the useful acquisition signal while substantially reducing the number of expensive utility evaluations. The benefit of this strategy is expected to become more important for larger candidate pools, where exhaustive enumeration becomes increasingly infeasible.

More advanced acquisition schedules, more carefully tuned surrogate models, or more advanced decision-making strategies may further reduce the evaluation cost or improve the overlap with full enumeration.

\subsection{Training setup}




Unless otherwise stated, we follow the default training setup in~\cite{zhang2024cut} for communication-structure optimization.

\subsection{Broader benchmark settings}\label{apd:broader_benchmark_settings}

\begin{table}[h]
\centering
\caption{Experimental settings for the broader benchmarks.}
\label{tab:broader_benchmark_setup}
\begin{small}
\begin{tabular}{lccccc}
\toprule
Benchmark & Selection pipeline & Agents / adv. & Test tasks & Metric & $l^*$ \\
\midrule
HumanEval
& $94 \rightarrow 50 \rightarrow 10$
& $5/3$ & 50 
& Accuracy & 1 \\

InterCode-SQL
& $834 \rightarrow 50 \rightarrow 10$
& $5/1$ & 50 
& Success@5, AUC@5 & 1 \\

ALFWorld
& $20 \rightarrow 12 \rightarrow 10$
& $4/2$ & 20 
& Success@30, AUC@30 & 1 \\
\bottomrule
\end{tabular}
\end{small}
\end{table}

The selection pipeline reports the candidate-pool size, representative-pool size, and final number of selected training tasks, respectively. Each selected task is used twice during graph optimization.

For HumanEval, we use the binary code-execution as performance score for graph optimization, with $l=1$ if the generated code passes execution and $l=0$ otherwise, and set $l^*=1$.
For InterCode-SQL and ALFWorld, we combine task completion, early success, and intermediate progress as
\[
l =
0.6\,\mathrm{Success@}k
+0.2\,\mathrm{AUC@}k
+0.2\,\mathrm{BestReward@}k,
\]
where $k=5$ for InterCode-SQL and $k=30$ for ALFWorld. Success@$k$ measures the fraction of tasks solved within $k$ interaction steps, while AUC@$k$ averages success over the interaction horizon and therefore also reflects how early a task is solved. $\mathrm{BestReward@}k$ denotes the highest environment reward obtained within the first $k$ interaction steps. Since the maximum possible score is $1$, we set $l^*=1$ for both benchmarks.

\section{MAS setup details}\label{apd:mas_setup}

\subsection{General setup}

We directly adopt the MAS execution framework of AgentPrune~\cite{zhang2024cut} in all experiments. Although the present experiments are built on this specific system, the proposed active learning method is largely independent of the particular MAS construction, since it mainly treats the MAS as a black-box forward model under different communication graphs. Below we summarize the execution details most relevant to our experiments.

The multi-agent system is represented by a spatial-temporal communication graph, where the spatial graph describes information exchange within each interaction round and the temporal graph describes how information is passed across adjacent rounds. Each node corresponds to an agent. In our setting, the graph structure is parameterized by trainable logits, represented in our notation by the \(\mathbf{z}\). These logits are first transformed by a sigmoid function into edge-retention probabilities in \([0,1]\). A Bernoulli variable is then sampled for each edge to determine whether that communication edge is active in the current graph realization. In this way, the logits do not directly define a deterministic graph; rather, they parameterize a distribution over communication graphs.

Within each interaction round, agent execution must follow a valid order. Therefore, the spatial communication graph is executed as a directed acyclic graph (DAG), so that agents can be processed in topological order. Intuitively, an agent can only read messages that have already been produced by its spatial predecessors in the same round. The temporal graph, in contrast, determines which outputs from the previous round are passed forward to the current round. Thus, spatial edges control same-round communication, while temporal edges control cross-round memory and dialogue-history propagation.

Given an input task \(q\), the system runs for multiple rounds. At round \(t\), each agent receives the task together with messages from its temporal in-neighbors from round \(t-1\) and its spatial in-neighbors from the current round \(t\). The agent then produces its response conditioned on this collected context, along with its predefined role and internal state. Therefore, the communication graph affects the MAS entirely through message routing: it determines which agent outputs are visible to which other agents, both within a round and across rounds.

After the final round, a decision agent reads the resulting dialogue trajectory and produces the final answer. In our experiments, this entire execution pipeline, including the prompt family and role assignment, is inherited from AgentPrune; our contribution does not modify the MAS execution mechanism itself, but instead studies how to select informative tasks for optimizing its communication structure under limited training budgets.

\subsection{Agent roles, prompts, and attack details}\label{apd: roles}

\subsubsection{MMLU}

In this dataset, agent roles, prompts and attack details follow AgentPrune. We briefly introduce the main settings here. 

Agents are assigned role-specific prompts from the following role set: \textbf{Knowledgeable Expert, Critic, Mathematician, Psychologist, Historian, Doctor, Lawyer, Economist}, and \textbf{Programmer}. These role prompts are used to induce diverse reasoning perspectives across agents on the same multiple-choice question. More specifically, Knowledgeable Expert is prompted to identify key entities relevant to the question and mark them using the @...@ format. Critic is prompted to identify potential issues in other agents’ analyses. The remaining roles are defined through domain-oriented descriptions: Mathematician emphasizes mathematical reasoning and calculation; Psychologist emphasizes psychology, sociology, and philosophy; Historian emphasizes historical, political, economic, and social analysis; Doctor emphasizes medical reasoning; Lawyer emphasizes law, politics, and history; Economist emphasizes economics, finance, business, and chart interpretation; and Programmer emphasizes computer science, engineering, and physics.

All answering agents are given the same answer-format constraint. Each agent receives a four-option multiple-choice question with exactly one correct answer and is required to output one option from {A, B, C, D}. The response must be fewer than 100 words, with the first line containing only the selected option letter, followed by a brief step-by-step analysis. Agents are allowed to refer to other agents’ responses, but the analysis prompt explicitly instructs them to use such information only as additional advice and not to imitate other agents’ reasoning processes.

Answering agents are assigned roles by cycling through a fixed role pool. In the agent-attack setting, half of the agents are assigned the \textbf{Fake} role. The roles are allocated in an alternating manner: if one agent is assigned a regular role, the next agent is assigned Fake, and so on. The regular agents continue to cycle through the standard role pool described above.

For agent-attack experiments, adversarial agents are prompted with a dedicated attack template rather than the standard answering prompt. The attack prompt explicitly instructs the agent to provide an incorrect answer together with a false analysis process for the given question. It further specifies that, even if outputs from other agents are available, the adversarial agent should continue to produce deceptive content and try to mislead the other agents.

\subsubsection{GSM8K}

For GSM8K without agent attack, answering agents are assigned role-specific prompts from a fixed role pool consisting of \textbf{Math Solver, Mathematical Analyst, Programming Expert}, and \textbf{Inspector}. These roles are assigned in a cyclic manner across agents. The role prompts are designed to induce complementary solution styles for arithmetic reasoning: \textbf{Math Solver} is prompted to provide a direct step-by-step mathematical solution; \textbf{Mathematical Analyst} is prompted to first express the solution process in symbolic form and then substitute numerical values; \textbf{Programming Expert} is prompted to solve the problem using Python code only, with the final result assigned to the variable \verb|answer|; and \textbf{Inspector} is prompted to check the correctness of other agents’ reasoning or code and then provide its own step-by-step solution. In addition, the answer prompt prepends role-specific few-shot exemplars for \textbf{Math Solver}, \textbf{Mathematical Analyst}, and \textbf{Programming Expert} before the target question. 

The decision agent in GSM8K is prompted separately as a top decision-maker. It receives the mathematical problem together with analyses or code from other agents, and is instructed to provide the final answer by following the most common or most confident conclusion among the candidate outputs. Its response is required to end with a final line containing only the numerical result in the format “The answer is ...”, without units. 

In the agent-attack setting, some of the agents are replaced with \textbf{Fake} agents, while the remaining regular agents continue to use the standard GSM8K role pool described above. Fake agents are prompted to produce a convincing but incorrect step-by-step solution, to make the reasoning appear logical and confident, and to keep the final answer wrong. The prompt also explicitly requires the agent not to reveal that it is being misleading. They are required to follow or modify the incorrect answer from previous dialogue, so that their incorrect answers become more convincing.

\subsubsection{Clarification for attack agents}

In our current adversarial experiments, the adversarial agents are pre-specified in each experiment run and remain fixed across tasks and dialogue rounds within that run. We do not provide the identity of the adversarial agents to the other agents. Evaluating settings where the adversarial identity varies across tasks or even dialogue rounds is an important direction for future study.

\section{Additional results}

\subsection{Scalability to larger MAS and generalization across LLM backbones}
\label{apd:scale_backbone}

\subsubsection{Scalability}\label{apd:scalebili}

To evaluate whether the proposed method remains effective beyond the system scale considered in the main experiments (6 agents and 6 ensemble members), we further test adversarial MMLU with 10 and 20 agents. One-third of the agents are adversarial, and the EKI ensemble sizes are 10 and 15, respectively. 

\begin{table}[h]
\caption{Performance comparison across agent scales under agent attacks. Accuracy is reported as mean $\pm$ std, first quartile (Q1), and mean below Q1. Token cost is reported as mean.}
\label{tab:scale_backbone}
\vskip 0.15in
\begin{center}
\begin{small}
\begin{sc}\begin{tabular}{@{}lccllccc}
\toprule
& & \multicolumn{3}{c}{Accuracy / (\%)} & \multicolumn{3}{c}{Token cost / (M)} \\
\cmidrule(lr){3-5} \cmidrule(lr){6-8}
Case & Setup & Mean $\pm$ std & Q1 & Worst-25\% & Select & Train & Test \\
\midrule
\multirow{3}{*}{\shortstack{MMLU\\ 10 agents}}
& No train & 76.39 $\pm$ 2.26 & 74.83 & 73.62 & - & - & 2.40 \\
& Random & 77.27 $\pm$ 2.64 & 75.81 & 74.51 & - & 0.20 & 1.20 \\
& Active & 78.13 $\pm$ 1.85 & \textbf{76.80} (+0.99) & \textbf{75.61} (+1.10) & 2.80 & 0.20 & 1.20 \\
\midrule
\multirow{3}{*}{\shortstack{MMLU\\ 20 agents}}
& No train & 79.55 $\pm$ 2.36 & 77.12 & 76.47 & - & - & 7.40 \\
& Random & 78.43 $\pm$ 2.41 & 76.47 & 75.82 & - & 0.60 & 3.40 \\
& Active & 80.91 $\pm$ 2.17 & \textbf{79.73} (+3.26) & \textbf{78.75} (+2.93) & 13.00 & 0.60 & 3.60 \\
\bottomrule
\end{tabular}
\end{sc}
\end{small}
\end{center}
\vskip -0.1in
\footnotesize{Run counts for no training/random/active are 6/13/11 for 10-agent MMLU and 7/5/5 for 20-agent MMLU.}
\end{table}

Active selection outperforms random training at both larger MMLU scales, improving mean accuracy from 77.27\% to 78.13\% with 10 agents and from 78.43\% to 80.91\% with 20 agents. These results show that the benefit of active task selection is not restricted to the original six-agent system and can also be observed in larger systems.

For larger systems beyond the evaluated scale, EKI itself is not the primary computational bottleneck; the dominant scaling comes from the increased cost of a single MAS rollout itself. For a system with $N$ agents, the number of graph parameters scales as $O(N^2)$, while the token cost of one MAS rollout scales from $O(N)$ for sparse communication to $O(N^2)$ for dense communication. If the EKI ensemble size grows approximately linearly with $N$, the overall active-selection cost therefore scales between $O(N^2)$ and $O(N^3)$. Empirically, the selection cost increases from 0.8M tokens for 6 agents to 2.8M for 10 agents and 13.0M for 20 agents, approximately following quadratic growth over the tested range. The $O(N)$--$O(N^2)$ per-rollout cost is intrinsic to executing the MAS, while EKI contributes only the additional ensemble-size factor. For substantially larger systems, techniques such as localization or graph/ensemble dropout could further reduce the required ensemble size.

\subsubsection{Generalization across LLM backbones}\label{apd:llm_backbones}

We also evaluate HumanEval with a different LLM backbone to examine whether the observed benefit depends on the underlying LLM. We compare GPT-4.1-nano with GPT-5-nano and use five agents with three adversarial agents.

On HumanEval, active selection improves mean accuracy from 77.80\% to 78.80\% with GPT-4.1-nano and from 85.60\% to 86.40\% with GPT-5-nano. These results show that our method and its benefits are not limited to any specific LLM backbone.
 
\begin{table}[h]
\caption{Performance comparison across LLM backbones under agent attacks. Accuracy is reported as mean $\pm$ std, first quartile (Q1), and mean below Q1. Token cost is reported as mean.}
\label{tab:scale_backbone}
\vskip 0.15in
\begin{center}
\begin{small}
\begin{sc}\begin{tabular}{@{}cccccccc}
\toprule
& & \multicolumn{3}{c}{Accuracy / (\%)} & \multicolumn{3}{c}{Token cost / (M)} \\
\cmidrule(lr){3-5} \cmidrule(lr){6-8}
Case & Setup & Mean $\pm$ std & Q1 & Worst-25\% & Select & Train & Test \\
\midrule
\multirow{3}{*}{\shortstack{Human-\\Eval\\4.1-nano}}
& No train & 75.20 $\pm$ 4.44 & 73.00 & 69.33 & - & - & 0.23 \\
& Random & 77.80 $\pm$ 3.94 & 76.00 & 74.00 & - & 0.07 & 0.17 \\
& Active & 78.80 $\pm$ 4.44 & 76.00 & 74.67 & 0.64 & 0.09 & 0.17 \\
\midrule
\multirow{3}{*}{\shortstack{Human-\\Eval\\5-nano}}
& No train & 85.80 $\pm$ 2.20 & 84.50 & 83.33 & - & - & 0.28 \\
& Random & 85.60 $\pm$ 3.50 & 82.50 & 82.00 & - & 0.18 & 0.20 \\
& Active & 86.40 $\pm$ 3.98 & 84.00 & 82.67 & 0.80 & 0.23 & 0.20 \\
\bottomrule
\end{tabular}
\end{sc}
\end{small}
\end{center}
\vskip -0.05in
\footnotesize{Run counts for no training/random/active are 10/10/10 for both HumanEval configurations.}
\end{table}

\subsection{Additional results for adversarial setting}\label{apd: additional results}

In this section, we provide additional results for the adversarial setting. We include a representative example of graph change before and after optimization. We also report the context, EKI statistics, and simple comments on some of the tasks from the representative pool, in order to give an intuitive idea of what the tasks are, what kinds of tasks are informative, and how they affect the EKI process.

\subsubsection{A Representative Graph Change on MMLU}

\begin{figure}[t]
    \centering
    \includegraphics[width=\linewidth]{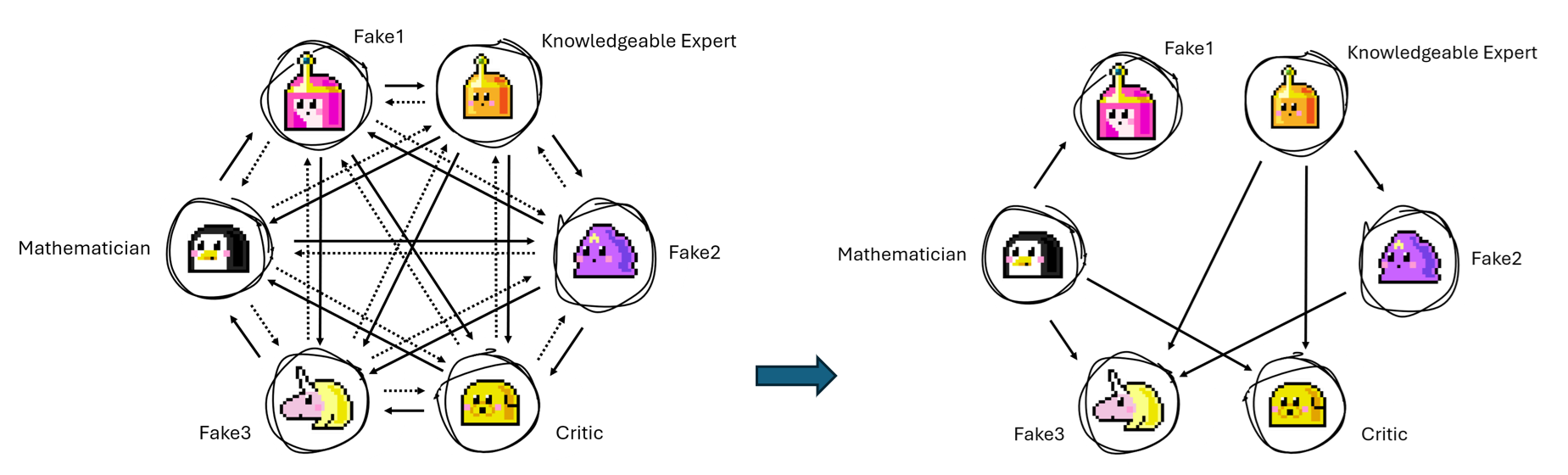}
    \caption{A representative example of graph change on MMLU under agent attack. The left panel shows the initial fully connected communication graph at the parameter level, while the right panel shows the optimized executable DAG after graph optimization. The learned structure suppresses several fake-to-normal communication paths while preserving a workable sparse topology.}
    \label{fig: change graph}
\end{figure}

Figure~\ref{fig: change graph} illustrates the practical effect of graph optimization in an adversarial setting. The left panel shows the initial communication graph as a fully connected topology at the parameter level, while the right panel shows the executable DAG after optimization. The optimized structure exhibits a clear pattern: normal agents no longer read messages from fake agents, indicating that the optimization process has learned to suppress harmful information flow from adversarial nodes. This behavior is consistent with the intended role of structure optimization under agent attack.

At the same time, the optimized DAG is not maximally sparse. Some fake agents still retain incoming edges from other normal or fake agents. We emphasize, however, that graph optimization is performed under a limited training budget rather than until full convergence. Under such a lightweight training setup, the learned graph already captures the main meaningful trend, namely, blocking fake-to-normal information flow while preserving a workable communication structure among useful agents.

\subsubsection{Selected tasks}

We show the details of several tasks to provide an intuitive understanding of the features of different tasks. We enumerate 50 tasks and evaluate information gain (KL-based) for each of them. We then rank these 50 tasks by their information gain. We first show the content of a few tasks in Fig.~\ref{fig:eki_task_examples}. We then show their corresponding information gain and other statistics within the EKI process in Table~\ref{tab:eki_rank_diagnostic}. We finally give a brief analysis of the difficulty of these tasks in Table~\ref{tab:eki_task_qualitative}, which is generated by ChatGPT-5.4.

In Table~\ref{tab:eki_rank_diagnostic}, \emph{Pred. var.} denotes the predictive variance across the ensemble, which reflects how differently the ensemble members respond to the same task. \emph{Cosine} measures the directional alignment between the ensemble before and after the update; a value close to one indicates that the update preserves a highly similar direction, while a smaller value suggests a more noticeable directional change. \emph{Var. sum} summarizes the overall ensemble variance after the update. These quantities provide complementary views of how each task interacts with the ensemble during the EKI update. Overall, a task with the largest information gain is often one for which the LLM-MAS exhibits high uncertainty. However, high uncertainty alone does not necessarily imply high informativeness, since the direction of the posterior update also matters.

The diagnostics show that tasks with large information gain tend to have large predictive variance and a smaller posterior ensemble variance. This suggests that the LLM-MAS is uncertain about these tasks, and that the corresponding EKI update substantially contracts the posterior ensemble. However, predictive uncertainty alone is not sufficient to determine informativeness. For example, the task at KL rank 16 still has a relatively large predictive variance, but its update direction remains close to that of the original ensemble and its posterior variance remains relatively large. This indicates that although the task induces substantial disagreement in the predicted outputs, this disagreement does not translate into a strong contraction of the graph-parameter posterior. Therefore, information gain depends not only on the magnitude of predictive uncertainty, but also on whether the task induces an informative update direction and effectively reduces posterior uncertainty.

Table~\ref{tab:eki_task_qualitative} gives a qualitative characterization of the same examples. The column \emph{Fact.} indicates whether the task requires domain-specific factual knowledge. \emph{Multi-concept} describes whether the task involves combining multiple concepts, rules, or conditions. \emph{Fine-grained} measures whether the answer depends on distinguishing closely related alternatives. \emph{Direct recall} indicates whether the task can be answered mainly by recalling a standard definition, term, or fact. This qualitative analysis helps connect the numerical EKI diagnostics with visible task structures. The interpretation column is intended to summarize the overall requirement of each task, including the reasoning steps it requires, the source of challenge, and whether the answer is mainly obtained through rule application, fine-grained comparison, or direct recall. Overall, this table provides a rough indication of task difficulty.

\begin{center}
\captionof{figure}{
Representative MMLU tasks at different EKI ranking positions from the same 50-task representative pool. The examples are shown to illustrate visible task structure rather than intrinsic task difficulty.
}
\label{fig:eki_task_examples}
\end{center}

\begin{taskbox}{KL rank 1, task index 11106}{}
\textbf{Question.}
A state enacts a statute that prohibits ``anyone over 60 years of age to run for public office.'' 
A state senator has been in office for three terms and wishes to seek re-election. 
The senator, who is 61, brings suit challenging the constitutionality of the state statute. 
Which of the following best states the burden of persuasion?

\textbf{Choices.}
(A) Since a fundamental right is involved, the state must show the regulation is necessary to vindicate a compelling government interest.\\
(B) Since no fundamental right is involved, the petitioner must show the age restriction is not rationally related to a legitimate government interest.\\
(C) The state must show the age regulation substantially furthers an important government objective and does not impair the fundamental right to vote.\\
(D) The petitioner must show the statute violates due process by depriving her of the right to be a candidate.

\smallskip
\textbf{Correct answer:} B.
\end{taskbox}

\begin{taskbox}{KL rank 6, task index 12110}{}
\textbf{Question.}
Which of the following goods is likely to have the most elastic demand curve?

\smallskip
\textbf{Choices.}
(A) Demand for white Ford minivans\\
(B) Demand for automobiles\\
(C) Demand for Ford automobiles\\
(D) Demand for American-made automobiles

\smallskip
\textbf{Correct answer:} A.
\end{taskbox}

\begin{taskbox}{KL rank 11, task index 9510}{}
\textbf{Question.}
Mary, a wealthy St. Petersburg widow, executed her first and only will on May 15, 1990 and died on August 18, 1990. 
Her will provided that her estate be divided equally between her only child, Joan, and the Salvation Army of Largo. 
How will Mary's estate actually be distributed?

\smallskip
\textbf{Choices.}
(A) 100\% to Joan.\\
(B) 100\% to Joan if she files a timely petition requesting that the devise to the Salvation Army be avoided.\\
(C) 50\% to Joan and 50\% to the Salvation Army.\\
(D) 50\% to Joan and the income from the remaining 50\% to Joan for life, remainder to the Salvation Army, if Joan files a timely petition protesting the devise to the Salvation Army.

\smallskip
\textbf{Correct answer:} C.
\end{taskbox}

\begin{taskbox}{KL rank 16, task index 8221}{}
\textbf{Question.}
Which of the following could result in autocorrelated residuals?

\smallskip
(i) Slowness of response of the dependent variable to changes in the values of the independent variables

(ii) Over-reactions of the dependent variable to changes in the independent variables

(iii) Omission of relevant explanatory variables that are autocorrelated

(iv) Outliers in the data

\smallskip
\textbf{Choices.}
(A) (ii) and (iv) only\\
(B) (i) and (iii) only\\
(C) (i), (ii), and (iii) only\\
(D) (i), (ii), (iii), and (iv)

\smallskip
\textbf{Correct answer:} C.
\end{taskbox}

\begin{taskbox}{KL rank 21, task index 1323}{}
\textbf{Question.}
In Ward-Leonard system, the lower limit of the speed imposed by

\smallskip
\textbf{Choices.}
(A) Field resistance.\\
(B) Armature resistance.\\
(C) Residual magnetism of the generator.\\
(D) None of above.

\smallskip
\textbf{Correct answer:} C.
\end{taskbox}

\begin{taskbox}{KL rank 26, task index 1918}{}
\textbf{Question.}
Which of the following anatomical regions of abdomen lies just distal to the sternum?

\smallskip
\textbf{Choices.}
(A) Epigastric\\
(B) Hypochondriac\\
(C) Hypogastric\\
(D) Lumbar

\smallskip
\textbf{Correct answer:} A.
\end{taskbox}

\begin{taskbox}{KL rank 31, task index 10592}{}
\textbf{Question.}
The vibrations in a transverse wave move in a direction

\smallskip
\textbf{Choices.}
(A) along the wave\\
(B) perpendicular to the wave\\
(C) Both of these\\
(D) Neither of these

\smallskip
\textbf{Correct answer:} B.
\end{taskbox}

\begin{taskbox}{KL rank 46, task index 12025}{}
\textbf{Question.}
Max was typically out of control whenever he attended preschool. 
Teachers tried time-outs and other punishments to no avail. 
His parents and the school decided to work with Max by giving him a sticker each time he behaved for a full hour. 
Once he accumulated ten stickers, he could present them to his parents who would give him a reward. 
The method the school and parents chose to employ is referred to as

\smallskip
\textbf{Choices.}
(A) negative reinforcement\\
(B) a token economy\\
(C) a point value system\\
(D) negative punishment

\smallskip
\textbf{Correct answer:} B.
\end{taskbox}

\begin{table}[H]
\caption{Diagnostic comparison of tasks at different EKI ranking positions.}
\label{tab:eki_rank_diagnostic}
\centering
\begin{small}
\begin{tabular}{rrrrrr}
\toprule
KL rank & Task index & KL & Pred. var. & Cosine & Var. sum \\
\midrule
1  & 11106 & 3804.28 & 86.32 & 0.056 & 131.36 \\
6  & 12110 & 91.31   & 49.52 & 0.526 & 121.54 \\
11 & 9510  & 29.08   & 54.70 & 0.822 & 162.18 \\
16 & 8221  & 12.44   & 76.60 & 0.862 & 206.03 \\
21 & 1323  & 3.70    & $9.125{\times}10^{-9}$  & 0.959 & 220.80 \\
26 & 1918  & 0.01    & $1.291{\times}10^{-7}$  & 1.000 & 257.92 \\
31 & 10592 & 0.00    & $3.706{\times}10^{-7}$  & 1.000 & 261.38 \\
46 & 12025 & 0.00    & $5.271{\times}10^{-9}$  & 1.000 & 261.09 \\
\bottomrule
\end{tabular}
\end{small}
\end{table}

\begin{table}[H]
\caption{Qualitative characterization of representative tasks at different EKI ranks. 
Scores use a 0--2 scale, where 0 means the property is not salient, 1 means moderately present, and 2 means clearly present.}
\label{tab:eki_task_qualitative}
\centering
\begin{small}
\begin{tabular}{rrrrrrp{0.44\linewidth}}
\toprule
\shortstack{KL\\ rank}  & \shortstack{Task\\ index} & Fact. & \shortstack{Multi-\\concept} & \shortstack{Fine-\\grained} & \shortstack{Direct\\ recall} & Task-structure interpretation \\
\midrule
1  & 11106 & 2 & 2 & 2 & 0 
& This legal reasoning task requires more than recognizing a constitutional-law term. The solver must identify the relevant legal category, determine whether candidacy is treated as a fundamental right, and then match that classification to the correct burden of persuasion. The main challenge is the multi-step rule selection and the distinction among closely related levels of scrutiny. \\

6  & 12110 & 2 & 1 & 2 & 1 
& This economics task tests a standard concept, but the answer depends on comparing several nested market definitions. The challenge is fine-grained rather than computational: the solver must recognize that demand becomes more elastic when the good is defined more narrowly and then distinguish the most specific option. \\

11 & 9510  & 2 & 2 & 2 & 0 
& This legal rule-application task combines factual details about timing, beneficiary type, and estate distribution. The solver must decide which legal rule applies to the given factual setting before selecting the distribution outcome. The challenge lies in applying a domain-specific rule to a concrete scenario rather than recalling an isolated fact. \\

16 & 8221  & 2 & 2 & 2 & 0 
& This econometrics task requires evaluating several proposed causes of autocorrelated residuals and then combining the valid statements. The challenge is that the answer cannot be obtained from a single keyword; each statement must be judged separately, and the final option depends on a correct aggregation of multiple conditions. \\

21 & 1323  & 2 & 0 & 1 & 2 
& This engineering task mainly depends on specialized factual knowledge about the Ward-Leonard system. The problem provides limited contextual information from which the answer can be inferred, so the challenge is primarily whether the solver has memorized or previously learned the relevant technical property. \\

26 & 1918  & 2 & 0 & 1 & 2 
& This anatomy task asks the solver to map a spatial description to the corresponding anatomical region. The main requirement is domain-specific terminology recall, with only a modest spatial distinction among the options. The challenge is therefore localized to recognizing the correct anatomical term. \\

31 & 10592 & 2 & 0 & 0 & 2 
& This physics task is close to a definition-matching question. Once the solver recalls the definition of a transverse wave, the answer follows directly. The options are also organized as broad opposites, which makes the required distinction relatively direct. \\

46 & 12025 & 2 & 1 & 1 & 1 
& This psychology task presents a concrete behavioral scenario and asks the solver to identify the corresponding concept. The challenge is to map the described reward-token procedure to the correct behavioral term, but the scenario contains strong cues that directly point to the answer. \\
\bottomrule
\end{tabular}
\end{small}
\end{table}


\subsection{Summary of statistical tests}
\label{apd:stat_summary}

Table~\ref{tab:stat_summary} summarizes the statistical tests for all headline comparisons between active learning and random training. For each benchmark, we test three statistics across independent runs: the mean, the first quartile (Q1), and the worst-25\% mean. Q1 and the worst-25\% mean measure the lower tail of performance. Appendices~\ref{apd:run_counts} and~\ref{apd:intercode_statistics} give more details.

We use two tests. On InterCode-SQL, the compared methods share the same random seeds, so we pair the runs by seed and use an exact paired permutation test. On MMLU, GSM8K, and HumanEval, we treat the runs of each method as independent samples and use a nonparametric bootstrap. For these three benchmarks, we report whether the 95\% confidence interval of the difference excludes zero.

We do not test significance on ALFWorld. Each method has only four to five runs on this benchmark. With five paired runs, the smallest attainable two-sided $p$-value of an exact paired permutation test is $2/2^5 = 0.0625$, and it is larger with four runs. No difference on ALFWorld can therefore be significant at the 0.05 level.

The mean improvement is significant on GSM8K with attack and on InterCode-SQL Success@5. The lower-tail improvement is significant on MMLU with attack (worst-25\% mean) and on GSM8K with attack (Q1 and worst-25\% mean). The other differences are not significant under the current number of runs.

\begin{table}[h]
\caption{Statistical tests for all headline comparisons between active learning and random training. For the bootstrap, we report whether the 95\% confidence interval of the difference excludes zero. For the paired permutation test, we report two-sided $p$-values. Bold entries are significant at the 0.05 level.}
\label{tab:stat_summary}
\vskip 0.15in
\centering
\begin{small}
\begin{tabular}{llccc}
\toprule
Benchmark & Test & Mean & Q1 & Worst-25\% mean \\
\midrule
MMLU, benign  & Bootstrap 95\% CI & includes 0 & includes 0 & includes 0 \\
GSM8K, benign & Bootstrap 95\% CI & includes 0 & includes 0 & includes 0 \\
MMLU, attack  & Bootstrap 95\% CI & includes 0 & includes 0 & \textbf{excludes 0} \\
GSM8K, attack & Bootstrap 95\% CI & \textbf{excludes 0} & \textbf{excludes 0} & \textbf{excludes 0} \\
HumanEval     & Bootstrap 95\% CI & includes 0 & includes 0 & includes 0 \\
\midrule
InterCode-SQL, Success@5 & Paired permutation & {\boldmath$p = 0.037$} & $p = 0.172$ & $p = 0.156$ \\
InterCode-SQL, AUC@5     & Paired permutation & $p = 0.066$ & $p = 0.133$ & $p = 0.156$ \\
\bottomrule
\end{tabular}
\end{small}
\vskip -0.1in
\end{table}

\subsubsection{Why some gains are smaller}

For some benchmarks (typically the easier ones), there is not much room for improvement. On these benchmarks, graph optimization with randomly selected tasks improves performance by only about 1 pp, similar to \citep{zhang2024cut}. Active selection can at most reach what graph optimization can achieve. Its gain over random selection comes from avoiding the poor task sets that lead to failure cases. This lifts the lower tail and, with it, the mean. So the mean gains on these benchmarks are expected to be small.

However, on the new benchmarks that we add and the original work \citep{zhang2024cut} does not use, i.e., InterCode-SQL and ALFWorld, the improvements are relatively large (typically 3–8 pp).

\subsubsection{Detailed statistics for broader benchmarks}\label{apd:broader_benchmark_statistics}

Table~\ref{tab:broader_benchmark_statistics} reports the detailed
performance statistics corresponding to the broader-benchmark results
in Fig.~\ref{fig:broader_benchmarks}.

\begin{table}[h]
\centering
\caption{Performance statistics for the broader benchmarks under adversarial
settings. HumanEval results correspond to GPT-4.1-nano. Performance is reported
as mean $\pm$ std, first quartile (Q1), and mean below Q1.}
\label{tab:broader_benchmark_statistics}
\begin{small}
\begin{tabular}{llccccc}
\toprule
Benchmark & Metric & Setup & Runs & Mean $\pm$ std & Q1 & Worst-25\% \\
\midrule

\multirow{2}{*}{HumanEval}
& \multirow{2}{*}{Accuracy}
& Random & 10 & $77.80 \pm 3.94$ & 76.00 & 74.00 \\
&
& Active & 10 & $78.80 \pm 4.44$ & 76.00 & 74.67 \\
\midrule

\multirow{4}{*}{InterCode-SQL}
& \multirow{2}{*}{Success@5}
& Random & 11 & $51.45 \pm 5.15$ & 50.00 & 45.33 \\
&
& Active & 11 & $57.27 \pm 4.22$ & 53.00 & 52.00 \\

& \multirow{2}{*}{AUC@5}
& Random & 11 & $28.84 \pm 2.91$ & 27.20 & 24.93 \\
&
& Active & 11 & $31.67 \pm 2.20$ & 30.40 & 28.93 \\
\midrule

\multirow{4}{*}{ALFWorld}
& \multirow{2}{*}{Success@30}
& Random & 5 & $45.00 \pm 17.80$ & 33.75 & 30.00 \\
&
& Active & 4 & $51.25 \pm 14.36$ & 48.75 & 30.00 \\

& \multirow{2}{*}{AUC@30}
& Random & 5 & $18.12 \pm 9.12$ & 13.17 & 8.17 \\
&
& Active & 4 & $20.88 \pm 5.61$ & 19.54 & 12.67 \\
\bottomrule
\end{tabular}
\end{small}
\end{table}

\subsubsection{Run counts and bootstrap confidence intervals of MMLU, GSM8K, and HumanEval}
\label{apd:run_counts}

Table~\ref{tab:run_counts} reports the number of independent runs used to compute the accuracy statistics in the main table (Table~\ref{tab:main_results}). Each run uses a different random seed. For random training, the random seed controls the selection of training tasks. For active learning, it controls the initial 1000-task pool realization and the stochastic components in the active-learning procedure.

Table~\ref{tab:bootstrap_ci} reports bootstrap 95\% confidence intervals for the main active-versus-random comparisons. The reported differences are computed as active learning minus random training for three metrics: mean accuracy, first quartile (Q1), and worst-25\% mean. These intervals quantify the uncertainty of the estimated differences across independent runs.

We compute the intervals using a nonparametric bootstrap. For each comparison, we resample runs with replacement within each method while keeping the original number of runs in that method. We then evaluate the target difference for each bootstrap sample. This procedure is repeated 10,000 times, and the 2.5th and 97.5th percentiles of the resulting bootstrap distribution are reported as the 95\% confidence interval.


Overall, the bootstrap intervals provide the strongest evidence under adversarial conditions. For GSM8K with attack, the 95\% confidence intervals for all three metrics exclude zero, while for MMLU with attack, the interval for worst-25\% mean excludes zero. The remaining intervals include zero. These results support particularly robust improvements in lower-tail performance under attack, while improvements whose intervals include zero are interpreted as observed effect-size trends rather than statistically decisive differences under the current number of independent runs.

\begin{table}[h]
\caption{Number of independent runs used for the accuracy statistics reported in the main tables.}
\label{tab:run_counts}
\vskip 0.15in
\centering
\begin{small}
\begin{tabular}{llccc}
\toprule
Setting & Dataset & No train & Random & Active \\
\midrule
Benign & MMLU  & 13 & 17 & 17 \\
Benign & GSM8K & 11 & 14 & 13 \\
Attack & MMLU  & 11 & 20 & 12 \\
Attack & GSM8K & 10 & 23& 23\\
Attack & HumanEval & 10 & 10 & 10 \\
\bottomrule
\end{tabular}
\end{small}
\vskip 0.15in
\end{table}

\begin{table}[h]
\caption{Bootstrap 95\% confidence intervals for EKI/active-learning improvements over random training. Differences are computed as EKI/active learning minus random training.}
\label{tab:bootstrap_ci}
\vskip 0.15in
\begin{center}
\begin{small}
\begin{tabular}{llccc}
\toprule
Dataset & Setting & Metric & Difference & 95\% bootstrap CI \\
\midrule
\multirow{3}{*}{MMLU}
& \multirow{3}{*}{Without attack} & Mean accuracy & +0.69 & $[-0.58, 1.91]$ \\
& & Q1 & +0.68 & $[-1.30, 2.64]$ \\
& & Worst-25\% & +1.30 & $[-0.12, 2.99]$ \\
\midrule
\multirow{3}{*}{GSM8K}
& \multirow{3}{*}{Without attack} & Mean accuracy & +0.28 & $[-0.14, 0.71]$ \\
& & Q1 & +0.32 & $[-0.11, 0.98]$ \\
& & Worst-25\% & +0.70 & $[-0.11, 0.92]$ \\
\midrule
\multirow{3}{*}{MMLU}
& \multirow{3}{*}{With attack} & Mean accuracy & +1.45 & $[-0.38, 3.34]$ \\
& & Q1 & +3.27 & $[-0.65, 5.23]$ \\
& & Worst-25\% & +4.25 & $[1.53, 6.32]$ \\
\midrule
\multirow{3}{*}{GSM8K}
& \multirow{3}{*}{With attack} & Mean accuracy & +0.93 & $[0.41, 1.45]$ \\
& & Q1 & +0.98 & $[0.33, 1.67]$ \\
& & Worst-25\% & +1.23 & $[0.72, 1.71]$ \\
\midrule
\multirow{3}{*}{HumanEval}
& \multirow{3}{*}{With attack} & Mean accuracy & +1.00 & $[-2.40, 4.60]$ \\
& & Q1 & +0.00 & $[-3.50, 4.00]$ \\
& & Worst-25\% & +0.67 & $[-2.67, 4.00]$ \\
\bottomrule
\end{tabular}
\end{small}
\end{center}
\vskip -0.1in
\end{table}

\subsubsection{Paired statistical analysis on InterCode-SQL}
\label{apd:intercode_statistics}

Unlike the MMLU and GSM8K experiments, the InterCode-SQL runs are paired by random seed; we therefore use a paired statistical test for this benchmark.

For InterCode-SQL, results are reported over 11 paired runs, with the compared methods matched by random seed. We therefore assess pairwise differences using a two-sided exact paired permutation test. Under the null hypothesis of no systematic difference between two methods, we enumerate all possible swaps of the two method labels within each paired run and compute the resulting distribution of the mean difference.

Table~\ref{tab:intercode_significance} reports the pairwise comparisons. Active learning improves Success@5 over random training by 5.82 percentage points ($p=0.037$), while the corresponding AUC@5 improvement is 2.84 percentage points ($p=0.066$). Relative to no training, active learning improves Success@5 by 9.45 percentage points ($p=0.093$) and AUC@5 by 6.29 percentage points ($p=0.047$).

Overall, these results provide additional statistical evidence that active task selection improves performance on InterCode-SQL, particularly in Success@5 relative to random training.

\begin{table}[h]
\caption{Two-sided exact paired permutation tests for InterCode-SQL over 11 paired runs. Differences are computed as the first method minus the second method.}
\label{tab:intercode_significance}
\vskip 0.15in
\centering
\begin{small}
\begin{tabular}{llcc}
\toprule
Metric & Comparison & Mean difference & $p$-value \\
\midrule
\multirow{3}{*}{Success@5}
& Random $-$ No train & +3.64 pp & 0.373 \\
& Active $-$ No train & +9.45 pp & 0.093 \\
& Active $-$ Random & +5.82 pp & 0.037 \\
\midrule
\multirow{3}{*}{AUC@5}
& Random $-$ No train & +3.45 pp & 0.094 \\
& Active $-$ No train & +6.29 pp & 0.047 \\
& Active $-$ Random & +2.84 pp & 0.066 \\
\bottomrule
\end{tabular}
\end{small}
\vskip -0.1in
\end{table}

\subsection{Sensitivity study}\label{apd:sensitivity}

\subsubsection{Ensemble size}\label{apd:ensemble_size}

We check the sensitivity of the task selection to ensemble size, using the MMLU dataset without agent attack. We fix the task pool of 50, select the top-10 tasks using different ensemble sizes and check the overlap ratio between the selected top-10 tasks and the reference top-10 tasks obtained with ensemble size 20. We repeat this experiment nine times with different initial ensembles and task pools, and error bars denote one standard deviation across repeated runs. For a baseline, a random top-10 ranking from the 50-task pool would have an expected overlap ratio of $10/50=0.2$ with the reference top-10 set. The results are shown as the $d=36$ curve in Fig.~\ref{fig:overlap_small_ens}.

As expected, the overlap ratio increases with ensemble size, indicating that larger ensembles lead to more stable task identification. At the same time, even a relatively small ensemble already recovers most of the tasks selected by the larger-ensemble reference. In particular, ensemble size 6 achieves an average overlap ratio of around 0.75. This does not mean that only 7.5 out of the 10 selected tasks are truly valuable. Some tasks ranked just below the top 10 under ensemble size 20 may still be informative and can be selected when using a smaller ensemble. Therefore, the overlap ratio should be interpreted as a measure of agreement with the reference selection, rather than a strict count of useful tasks. Overall, the results suggest that \textbf{EKI-based task selection remains effective even with a small ensemble}, while larger ensembles further improve stability. 

Note that we use ensemble size 20 as a larger-budget reference to study the stability trend, rather than to define a ground-truth ranking. Our goal here is not to fully characterize convergence with respect to ensemble size, but to examine whether a very small ensemble already provides effective task identification at substantially lower cost. This is consistent with prior observations in the Bayesian experimental design literature \cite{callahanreverse} that accurate inner-loop inference is not always necessary for useful EIG estimation, and that small inner sample sizes can substantially reduce computation while preserving practical utility.

\subsubsection{Diagonal versus full KL}
\label{apd:diag_vs_full}

Our method uses a diagonal KL, which ignores the correlations between parameters. To test this approximation, we compare it with the full-covariance KL in a reduced setting where the full covariance can be computed.

We keep the 36-parameter graph used in Appendix~\ref{apd:ensemble_size}. In each task pool, we randomly select four graph parameters and freeze the other 32, so the parameter dimension is $d=4$. We use three task pools of 50 tasks each from MMLU, and run three trials per pool with different random seeds. This gives nine runs. In each run, we perform the one-step EKI update with $J=3$, 6, 12, and 20, using independently drawn ensembles. For each candidate task, we compute the KL divergence between the updated and initial ensembles with two estimators:
\begin{itemize}
    \item \textbf{Diagonal KL}, used in our method. We keep only the diagonal of the ensemble covariance and set a lower bound of $10^{-8}$ on each variance.
    \item \textbf{Full KL.} We use the full $4\times 4$ ensemble covariance. For $J \geq 6$, the sample covariance is nonsingular, and we use it without regularization. For $J=3$, the sample covariance has rank at most 2. We add a ridge term $\lambda I_4$ to both the initial and the updated covariance, with $\lambda = 10^{-3}\,\mathrm{tr}(C_0)/4$, where $C_0$ is the initial ensemble covariance.
\end{itemize}
The diagonal KL needs $O(d)$ memory. The full KL needs $O(d^2)$ memory and $O(d^3)$ time for the inverse and the log-determinant of the covariance.

For each estimator and each $J$, we select the 10 tasks with the largest KL. We report two overlap ratios. Stability is the overlap between the Top-10 set at ensemble size $J$ and the Top-10 set at $J=20$ of the same estimator. Diag.\ vs.\ Full is the overlap between the Top-10 sets of the two estimators at the same $J$. To measure the run-to-run variation, we also compute the overlap between two $J=20$ runs with different seeds on the same task pool. Each pool has three such pairs, which gives nine pairs. A random selection of 10 tasks from 50 gives an expected overlap of $10/50=0.2$. Table~\ref{tab:diag_full} reports the results.

\textbf{The diagonal approximation has little effect on task selection.} At the same ensemble size, the two estimators share 8.1 to 9.1 of the 10 selected tasks on average (Table~\ref{tab:diag_full}, last column). Ignoring the parameter correlations therefore changes about one of the ten selected tasks. This supports the diagonal KL in our method, which also keeps the memory cost linear in $d$.

\textbf{Task selection with a small ensemble is as stable as with $J=20$.} For $J=3$ to 12, the stability stays between 0.67 and 0.74 for both estimators. Two $J=20$ runs with different seeds have a similar overlap: 0.70 for the diagonal KL and 0.73 for the full KL (last row). The difference between a small ensemble and $J=20$ is therefore comparable to the difference between two runs of $J=20$ itself. This variation comes from the initial ensemble, the graph sampling, and the LLM rollouts. The stability also does not increase steadily from $J=3$ to 12, and the changes are within one standard deviation. In the 36-parameter setting, the overlap at $J=6$ is around 0.75 (Appendix~\ref{apd:ensemble_size}), close to the value here. A rank-deficient covariance with $J<d$ is therefore not the main cause of the disagreement with $J=20$.

\begin{table}[h]
\caption{Top-10 overlap in the reduced setting with $d=4$. Stability is the overlap with the $J=20$ selection of the same estimator. Diag.\ vs.\ Full is the overlap between the two estimators at the same $J$. The last row compares two $J=20$ runs with different seeds on the same task pool. Values are mean $\pm$ standard deviation over nine runs (3 task pools $\times$ 3 seeds), or over nine pairs of runs for the last row. A random selection gives an expected overlap of 0.2.}
\label{tab:diag_full}
\vskip 0.15in
\centering
\begin{small}
\begin{tabular}{lccc}
\toprule
$J$ & Stability: Diag. & Stability: Full & Diag.\ vs.\ Full \\
\midrule
3  & $0.722 \pm 0.130$ & $0.733 \pm 0.071$ & $0.833 \pm 0.112$ \\
6  & $0.722 \pm 0.083$ & $0.711 \pm 0.105$ & $0.811 \pm 0.093$ \\
12 & $0.744 \pm 0.113$ & $0.667 \pm 0.087$ & $0.911 \pm 0.078$ \\
20 (as reference)     & $1.000$ & $1.000$ & $0.911 \pm 0.078$ \\
20 (vs.\ another run) & $0.700 \pm 0.071$ & $0.733 \pm 0.050$ & -- \\
\bottomrule
\end{tabular}
\end{small}
\vskip -0.1in
\end{table}

This test used GPT-4.1-mini, while the tests in \ref{apd:ensemble_size} and in \ref{apd:sensitivity_to_step} used GPT-4.1-nano. Because at the time we added this test, GPT-4.1-nano is no longer available.

\subsubsection{EKI iteration steps}\label{apd:sensitivity_to_step}

We further check the sensitivity of the task selection to the EKI iteration step, using the MMLU dataset without agent attack. We fix the task pool of 50, select the top-10 tasks using different EKI iteration steps and check the overlap ratio between the selected top-10 tasks and the reference top-10 tasks obtained with three EKI steps. We repeat this experiment nine times with different initial ensembles and task pools, and error bars denote one standard deviation across repeated runs. 

\begin{figure}[t]
    \centering
    \begin{minipage}[t]{0.39\linewidth}
        \centering
        \includegraphics[width=\linewidth]{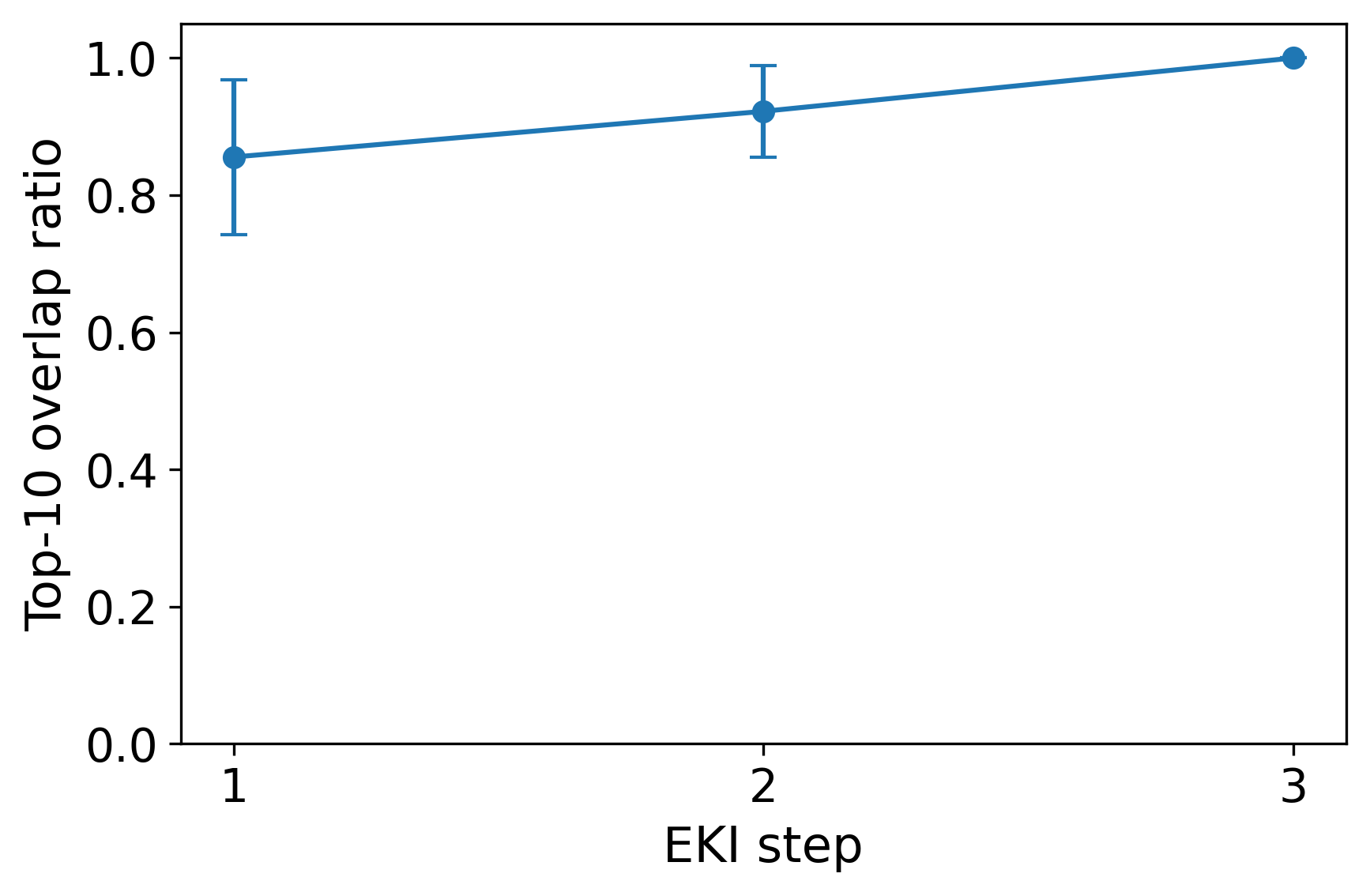}
        \captionof{figure}{Top-10 overlap with the final three-step ranking as a function of the number of EKI iterations.}
        \label{fig:sensitivity_step}
    \end{minipage}
    \hfill
    \begin{minipage}[t]{0.29\linewidth}
        \centering
        \includegraphics[width=\linewidth]{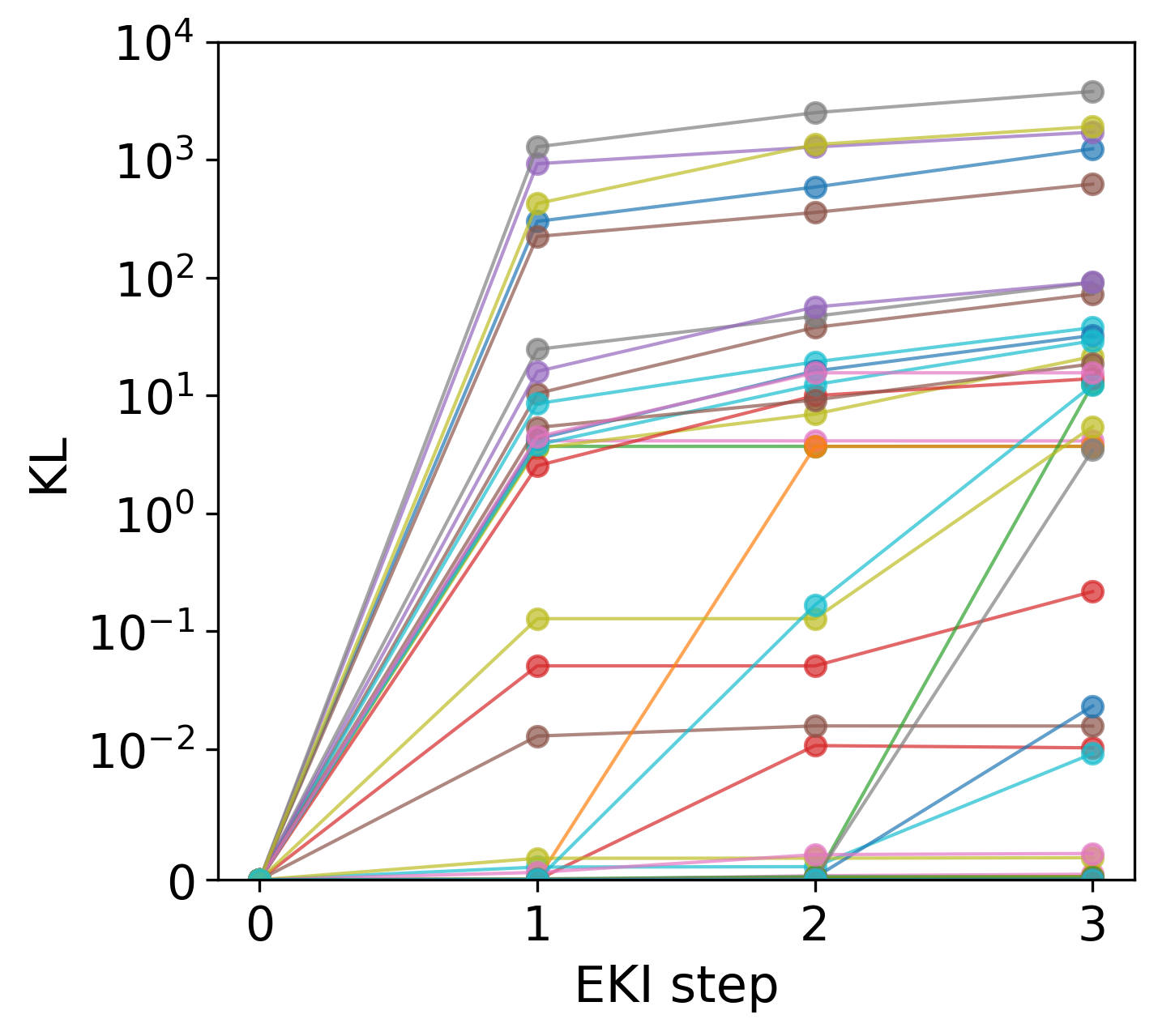}
        \captionof{figure}{KL trajectory over EKI iteration steps for one representative pool.}
        \label{fig:sensitivity_step_kl}
    \end{minipage}
    \hfill
    \begin{minipage}[t]{0.29\linewidth}
        \centering
        \includegraphics[width=\linewidth]{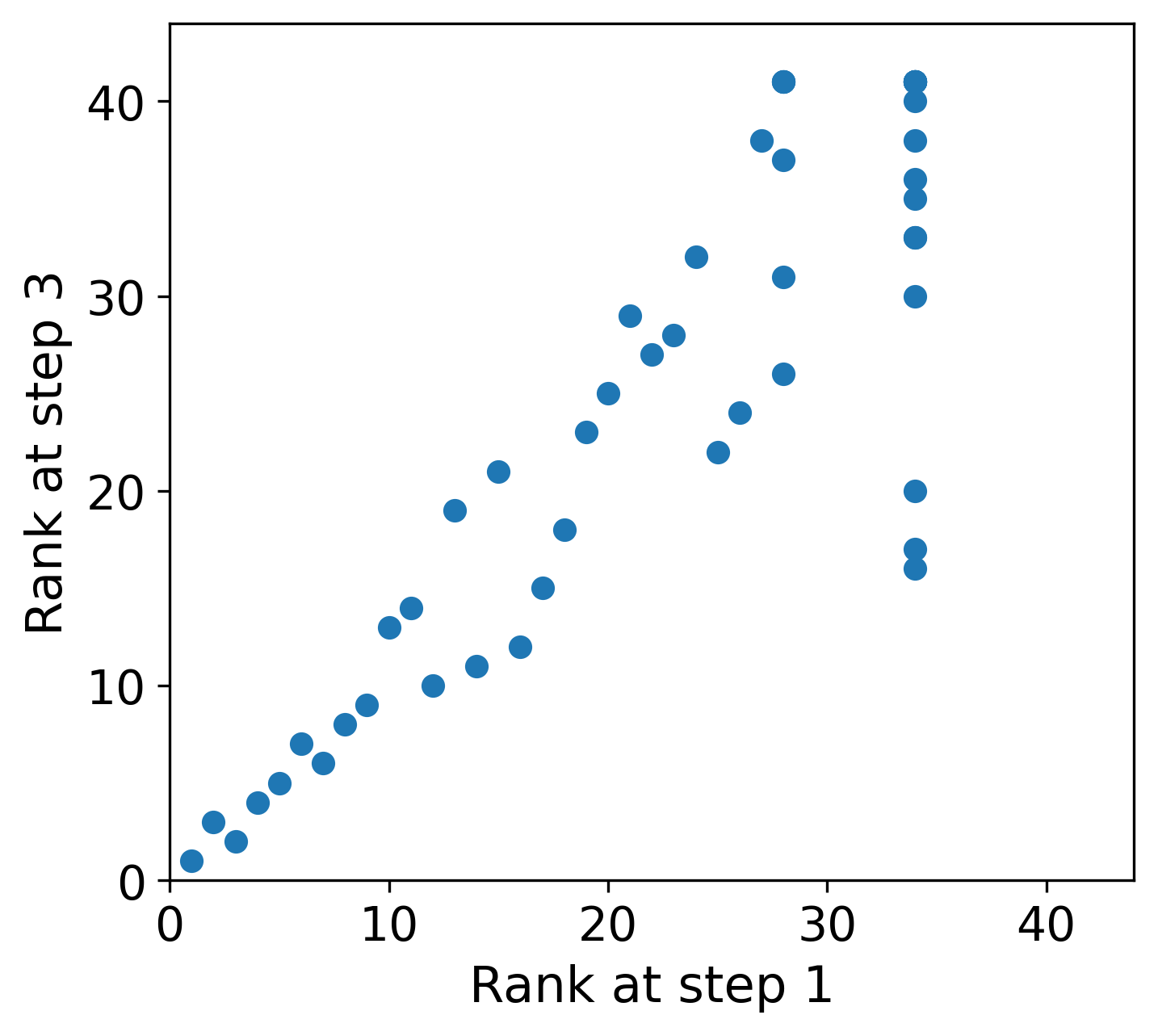}
        \captionof{figure}{Rank comparison between step 1 and step 3 for one representative pool.}
        \label{fig:sensitivity_step_scatter}
    \end{minipage}
\end{figure}

The overall results are shown in Fig.~\ref{fig:sensitivity_step}. The tasks selected by one-step EKI and three-step EKI exhibit a high degree of overlap: on average, 8--9 out of the top-10 selected tasks are the same. We further plot the KL trajectories over EKI steps for all 50 tasks in Fig.~\ref{fig:sensitivity_step_kl}. The top-ranked tasks maintain consistently larger KL values than the remaining tasks across EKI steps. Their rankings at step 1 and step 3 are also generally consistent, as shown in Fig.~\ref{fig:sensitivity_step_scatter}.

These results indicate that (i) \textbf{task selection is robust to the choice of the EKI iteration step}, and (ii) \textbf{one-step EKI is sufficient to identify the most informative tasks}.

\subsection{Comparison with other active learning methods}\label{apd: other methods}

In this section, we introduce how we implement other active learning methods and how we balance the selection cost. For EGL and EMC, the computational cost is largely determined by the number of forward samples used to approximate the expectation in the REINFORCE estimator. In our experiments, we choose this sampling budget such that the per-task gradient-based score estimation remains affordable, which makes direct enumeration over the 50-task pool feasible, without the need for Thompson sampling. For the Fisher-based methods, enumerating the same pool also yields a pool-level Fisher matrix, which enables a greedy subset selection strategy that partially accounts for batch interactions rather than selecting tasks independently.

\subsubsection{EKI total rollouts}

Based on the setup described in \ref{apd: ts}, we use EKI to measure information gain for 25 tasks out of 50 and choose the top-10. Since each task evaluation involves 6 ensemble members, the total number of rollouts is $6 \times 25=150$. We keep the number of rollouts for the other methods the same.

\subsubsection{EGL-based selection}

Expected gradient length quantifies task informativeness by the magnitude of the gradient, that is, by how strongly a task is expected to change the graph parameters. We use pointwise estimates rather than full expectations for the same reason as discussed in Section \ref{sec: eki}. In our setting, this gradient is the derivative of the communication-structure optimization utility with respect to the graph logits, estimated by policy gradient (REINFORCE). This is the same type of gradient used for graph optimization:
\begin{equation}
    h = \nabla_{\mathbf z}\,
    \mathbb{E}_{A}\,
    \mathbb{E}_{\{q_i,y_i\}\in\mathcal D_{\mathrm{tr}}}
    \big[\phi(g,q_i,y_i)\big].
\end{equation}
For task selection, we score each candidate task by the $\ell_2$ norm of its estimated gradient,
\begin{equation}
    U(q_i)=\|h_i\|_2,
\end{equation}
and then rank all candidate tasks by this score.

In the optimization implementation of AgentPrune, the expectation over $A$ is typically approximated using only one sample, which greatly reduces computational cost but also leads to a noisy gradient estimate. In our EGL implementation, we use $3$ rollout samples for each task and enumerate all $50$ tasks in the representative pool. Therefore, the selection uses a total of $150$ rollouts.

\subsubsection{EMC-based selection}

Expected model change quantifies task informativeness by the magnitude of the parameter change induced after several gradient-descent steps. For each candidate task, we start from the same initial graph, perform a short single-task training procedure, and measure how much the graph logits change before and after optimization.

Let $\mathbf z^{\mathrm{before}}$ and $\mathbf z^{\mathrm{after}}$ denote the flattened graph logits before and after the short training procedure, respectively. We define the task score as
\begin{equation}
    U(q_i)=\|\Delta \mathbf z_i\|_2,
    \qquad
    \Delta \mathbf z_i
    =
    \mathbf z^{\mathrm{after}}_i-\mathbf z^{\mathrm{before}}_i.
\end{equation}
We then rank all candidate tasks by this score and select the top ones.

In our implementation, we use one rollout sample in the REINFORCE gradient estimator for each update step, but run $3$ gradient-descent steps for each task. The optimizer is the same as in training, namely Adam with learning rate $0.1$.  The selection uses a total of $150$ rollouts.

\subsubsection{Fisher-based selection}

\textbf{Formula}. The Fisher-based method quantifies task informativeness through a rollout-level empirical Fisher matrix. Fisher information measures how sensitively the model response changes with respect to the parameters, and is widely used as a proxy for the amount of information that an observation provides about those parameters. In our setting, each rollout on a candidate task induces a gradient with respect to the graph logits, and the corresponding empirical Fisher matrix captures the magnitude and directional structure of these task-induced updates.

For each candidate task $q_i$, suppose we obtain $m$ rollout gradients
\(
h_{i,1}, h_{i,2}, \dots, h_{i,m}
\)
with respect to the graph logits. We construct the empirical Fisher matrix of this task as
\begin{equation}
    F_i
    =
    \frac{1}{m}\sum_{r=1}^{m} h_{i,r} h_{i,r}^{\top}.
\end{equation}

\textbf{Relationship to EGL}. It should be noted that when $m=1$, the task-wise Fisher matrix reduces to
\begin{equation}\label{eq:fisher_single}
    F_i = h_i h_i^\top,
\end{equation}
so that
\begin{equation}
    \mathrm{tr}(F_i) = \|h_i\|_2^2.
\end{equation}
Therefore, for single-rollout single-task scoring, the trace criterion is closely related to EGL, whose score is proportional to $\|h_i\|_2$, and differs only by a monotone transformation. In this sense, the two criteria induce the same ranking over tasks in the single-rollout case.

It should also be noted that, in this case, the determinant of the Fisher matrix is zero whenever the graph-parameter dimension is greater than one, since Eq.~\eqref{eq:fisher_single} is a rank-one matrix. As a result, the determinant of a single-task Fisher matrix does not provide a meaningful ranking criterion in the single-rollout setting.

With multiple rollouts, the empirical Fisher matrix becomes an average of gradient outer products across rollouts, which incorporates second-moment information beyond the averaged-gradient magnitude used in EGL. However, this trace-based or determinant-based score remains closely related in spirit to EGL, and in our view does not constitute a sufficiently distinct task-selection criterion to warrant a separate standalone comparison.

\textbf{Greedy subset selection}. To address this limitation, we do not score each task independently and then directly take the top-$k$. Instead, we perform greedy subset selection so as to construct a subset whose accumulated Fisher information better covers the representative pool. First, we build a baseline Fisher matrix by averaging the task-wise Fisher matrices over the representative pool:
\begin{equation}
    B_0
    =
    \frac{1}{N}\sum_{i=1}^{N} F_i + \lambda I,
\end{equation}
where $N=50$ is the representative-pool size and $\lambda$ is a small regularization constant.

Starting from $B=B_0$, at each greedy step we add one task that gives the largest marginal gain. In the trace-based version, the score of candidate task $q_i$ is
\begin{equation}
    U_{\mathrm{trace}}(q_i)
    =
    \mathrm{tr}(B^{-1}F_i).
\end{equation}
In the determinant-based version, the score is
\begin{equation}
    U_{\mathrm{det}}(q_i)
    =
    \log\det(B+F_i)-\log\det(B).
\end{equation}
After selecting a task, we update
\begin{equation}
    B \leftarrow B + F_i,
\end{equation}
and repeat this procedure until 10 tasks are selected.

We use 3 rollout samples per task and enumerate 50 tasks. This selection uses a total of $150$ rollouts.

The current formulation is closely related to A-optimality and D-optimality in Bayesian experimental design. For convenience, we refer to this baseline as \emph{Coreset} in the figures. More precisely, it is a coreset-style greedy subset-selection method based on task-wise empirical Fisher information, designed to select a compact subset that better covers the representative pool.

\subsubsection{Detailed statistics for Fig.~\ref{fig:other_al_methods}}\label{apd:detail_statistics_other_al}

Table~\ref{tab:baseline_comparison_attack} reports the numerical mean accuracy and standard deviation corresponding to the baseline comparison shown in Fig.~\ref{fig:other_al_methods}.

The Fisher coreset comparison should be interpreted under the fixed selection-cost budget used in our experiments. \textbf{We do not claim that task-wise EKI scoring generally dominates a fully optimized Fisher coreset method. Instead, the comparison reflects a practical cost trade-off}. Fisher coreset construction requires information from the entire 50-task pool, so under the same total selection cost, each task-level Fisher matrix is estimated with a limited rollout budget and may be noisy. In contrast, the EKI-based method avoids explicit pool-level coreset construction and directly estimates task-level utility. This may lead to a more reliable task-level informativeness estimate in the cost-limited black-box setting considered here.

Table~\ref{tab:baseline_bootstrap_ci} reports bootstrap 95\% confidence intervals for the auxiliary baseline comparison, using random training as the common reference. The results show that EKI and Fisher det have the largest observed improvements over random training, while EGL, EMC, and Fisher trace show smaller or negative observed differences. We interpret this comparison as descriptive evidence of the overall trend, rather than as a significance-ranked ordering among methods.

Between EKI and Fisher det, with the same bootstrap procedure, the 95\% confidence interval for their difference in mean accuracy (EKI minus Fisher det) is $[-2.17, 1.53]$ percentage points. \textbf{The interval includes zero, so the two methods are not statistically distinguishable in this experiment}. (The runs are not paired by random seed across methods, so a paired test is not applicable to the direct comparison )

Overall, these results suggest that EKI provides competitive performance in the cost-limited setting.

\begin{table}[t]
\caption{Numerical comparison of different task-selection baselines on MMLU under agent attack.}
\label{tab:baseline_comparison_attack}
\vskip 0.15in
\begin{center}
\begin{small}
\begin{sc}
\begin{tabular}{lcc}
\toprule
Method & Mean accuracy (\%) & Std. (\%) \\
\midrule
No train & 75.68 & 2.10 \\
Random train & 76.87 & 3.72 \\
Fisher coreset (trace) & 76.25 & 1.17 \\
EGL & 77.02 & 1.93 \\
EMC & 77.21 & 2.58 \\
EKI & 78.32 & 1.80 \\
Fisher coreset (determinant) & 78.59 & 1.91 \\
\bottomrule
\end{tabular}
\end{sc}
\end{small}
\end{center}
\vskip -0.1in
\end{table}

\begin{table}[h]
\caption{Bootstrap 95\% confidence intervals for mean-accuracy differences in the baseline comparison. Improvements are computed relative to random training.}
\label{tab:baseline_bootstrap_ci}
\vskip 0.15in
\begin{center}
\begin{small}
\begin{sc}
\begin{tabular}{lcc}
\toprule
Method & Improvement & 95\% bootstrap CI \\
\midrule
No train & -1.19 & $[-3.28, 0.83]$ \\
Fisher coreset (trace) & -0.62 & $[-2.46, 1.24]$ \\
EGL & +0.16 & $[-1.96, 2.34]$ \\
EMC & +0.34 & $[-1.96, 2.99]$ \\
EKI & +1.45 & $[-0.38, 3.34]$ \\
Fisher coreset (determinant) & +1.72 & $[-0.45, 3.99]$ \\
\bottomrule
\end{tabular}
\end{sc}
\end{small}
\end{center}
\vskip -0.1in
\end{table}

\subsubsection{Qualitative comparison between EKI and Fisher-determinant selection}
\label{apd:eki_fisher_examples}

Although EKI and Fisher-determinant selection achieve similar downstream
performance, they characterize task informativeness differently. EKI measures
the realized finite change in the graph-parameter distribution induced by a
task. In contrast, the Fisher matrix describes local parameter sensitivity, and
the Fisher-determinant coreset further accounts for redundancy and coverage
across the candidate pool. As a result, the two methods can assign substantially
different priorities to individual tasks.

We illustrate this difference using four representative MMLU tasks:

\begin{itemize}[leftmargin=*]
    \item \textbf{Task 13719 [virology\_test]:}
    How is the parvovirus family targeted to reduce disease?
    Correct answer: \textit{screen transfusion blood}.

    \item \textbf{Task 12110 [high\_school\_microeconomics\_test]:}
    Which good likely has the most elastic demand curve?
    Correct answer: \textit{white Ford minivans}.

    \item \textbf{Task 8221 [econometrics\_test]:}
    Which factors could cause autocorrelated residuals
    (slow response, over-reaction, omitted autocorrelated variables, or outliers)?
    Correct answer: \textit{i, ii, iii}.

    \item \textbf{Task 11587 [high\_school\_world\_history\_test]:}
    A question based on a passage from Felipe Guaman Poma de Ayala,
    \textit{The First New Chronicle and Good Government} (ca.~1610), asking
    which change in Spanish policy toward Native Americans resulted.
    Correct answer: \textit{royal decrees for more humane treatment}.
\end{itemize}

\begin{table}[h]
\centering
\caption{Representative task-level diagnostics comparing EKI-based information
gain and Fisher-determinant coreset selection. KL rank is the rank according to
the EKI-based information gain.}
\label{tab:eki_fisher_examples}
\vskip 0.1in
\begin{small}
\begin{tabular}{lccccc}
\toprule
Task index & KL rank & KL value & Prediction variance & Cosine &
Fisher-det rank \\
\midrule
13719 & 2  & 13.31 & 30.00       & 0.56 & 44 \\
12110 & 9  & 3.82  & 72.52       & 0.81 & 12 \\
8221  & 28 & 1.91  & 53.99       & 0.94 & 1  \\
11587 & 40 & 0     & $7.68\times10^{-10}$ & 1.00 & 35 \\
\bottomrule
\end{tabular}
\parbox{0.83\linewidth}{\footnotesize\textit{Note:} The values differ from those in Table~\ref{tab:eki_rank_diagnostic} because the two tables report results from different random-seed runs.}

\end{small}
\end{table}

The examples show several distinct cases. Task 13719 is ranked second by EKI
but only 44th by the Fisher-determinant coreset. Its relatively low
update-direction cosine (0.56) indicates a distinctive parameter update,
resulting in a large change in the parameter distribution and therefore
a high EKI information gain. In contrast, Task 8221 is ranked first by the
Fisher-determinant coreset but only 28th by EKI. This illustrates the
pool-level nature of coreset selection: a task can be valuable because its
local information complements that of other tasks even when its individual
EKI information gain is moderate.

Task 12110 is ranked highly by both methods, showing that the two criteria can
also agree on informative tasks. Task 11587 provides the opposite example:
its negligible predictive variation produces essentially zero EKI information
gain, and neither method assigns it a high priority. Overall, these examples
show that EKI emphasizes task-induced finite changes in the graph-parameter
distribution, whereas the Fisher-determinant coreset additionally emphasizes
complementary coverage of the candidate pool.

\subsubsection{When each method works better}
\label{apd:eki_fisher_when}

Which method is more suitable depends mainly on whether gradients can be estimated accurately and on the available memory.

\textbf{Gradient}. Fisher-based selection requires gradients with respect to the graph parameters, whereas EKI only requires forward evaluations. Fisher-based selection can work better when gradients can be estimated accurately, for example, with more rollouts per task than in our cost-limited setting or with a differentiable forward model. EKI is more suitable when gradients are unavailable or noisy.

\textbf{Memory}. For a dense $N$-agent graph with $O(N^2)$ graph parameters, explicitly storing a dense Fisher matrix requires $O(N^4)$ memory, whereas EKI stores an ensemble with memory $O(JN^2)$. EKI is therefore more suitable when storing a dense Fisher matrix becomes expensive. 


\subsection{Ablation studies}\label{apd:ablation}

We conduct an ablation study on MMLU under the adversarial setting to isolate the contributions of the two key components in our active learning pipeline: representative selection and informative selection. The full method first selects a 50-task representative pool from the 1000-task candidate set, and then chooses 10 tasks from this pool using the proposed informativeness criterion. To understand the role of each component, we remove one component at a time while keeping the overall training budget and evaluation protocol unchanged. Specifically, ``Representative'' keeps the representative-selection stage but replaces informative selection with random choice from the representative pool. ``Informative'' replaces the representative pool with a random 50-task subset from the 1000-task candidate set, and then applies informative selection to choose 10 tasks. ``Representative + Informative'' denotes the full method.

As shown in Table~\ref{tab:ablation}, the three ablated variants exhibit a clear pattern. Representative selection alone does not lead to a consistent improvement over the baselines, while informative selection alone already provides a noticeable gain. The full method, which combines representative and informative selection, achieves the best mean accuracy and the smallest variability across runs.

These results indicate that informative selection is the primary source of performance improvement. This is because the informative stage directly selects tasks that are estimated to be beneficial for the downstream objective from the 50-task pool. As long as the pool contains sufficiently useful tasks, informative selection can extract them even when the pool itself is constructed randomly, which helps explain the improved minimum accuracy of the ``Informative'' setting. At the same time, without representative selection, the 50-task pool may provide weaker coverage of the broader candidate set and may contain more tasks with overlapping update directions, which can limit the strongest downstream outcomes.

While informative selection accounts for most of the gain, representative selection still provides a measurable additional benefit when combined with it, and is crucial for computational tractability.  Its role is not only to improve accuracy, but also to make the informative stage practical. Compared with using a random 50-task subset before informative selection, using representative selection to construct the 50-task pool leads to better final performance. This shows that the representative stage does not merely reduce the pool size, but meaningfully improves the quality of the candidate tasks passed to the informative stage. In addition, this reduction is crucial for efficiency: applying informative selection directly to the full candidate pool would be computationally impractical. The representative stage thus plays a dual role, serving both as a quality-improving filter and as an efficiency-enabling step for the subsequent informative selection.

\begin{table}[h]
\caption{Ablation study on different training setups. Accuracy is reported as mean $\pm$ std, together with the minimum and maximum values across runs.}
\label{tab:ablation}
\vskip 0.15in
\begin{center}
\begin{small}
\begin{sc}
\begin{tabular}{llccc}
\toprule
Setup & Variant & Accuracy / (\%) & Min & Max \\
\midrule
\multicolumn{2}{l}{No train} 
& 75.68 $\pm$ 2.10 & 70.58 & 79.08 \\
\multicolumn{2}{l}{Random train} 
& 76.87 $\pm$ 3.72 & 68.62 & 82.35 \\
\multirow{3}{*}{Active learning}
  & Representative selection only& 75.74 $\pm$ 3.64 & 70.58 & 81.04 \\
  & Informative selection only& 77.12 $\pm$ 2.45 & 72.54 & 79.73 \\
  & Representative + Informative & 78.32 $\pm$ 1.80 & 75.81 & 82.35 \\
\bottomrule
\end{tabular}
\end{sc}
\end{small}
\end{center}
\vskip -0.1in
\end{table}

\section{Scope and generalizability}\label{apd:scope_generalizability}

\subsection{Failure modes of information-gain-based task selection}\label{apd:failure_modes}

To address potential concerns about whether high information gain translates into downstream utility, we identify three possible failure modes and discuss how they are mitigated in our framework. First, an atypical or outlier task may exhibit high information gain without being broadly useful. Our representative-selection stage reduces this risk by restricting informative selection to tasks that represent non-negligible regions of the task distribution. Second, different task groups may favor conflicting communication structures. Our formulation follows the shared-graph setting of the adopted graph optimizer and therefore assumes that the task pool is not so heterogeneous that different groups require fundamentally different graphs. When distinct task groups do require different communication structures, the same task-selection framework can instead be applied within each group to identify informative tasks for group-specific graph optimization. Third, noisy evaluations may spuriously inflate the estimated information gain. This is primarily an estimation issue, for which the ensemble-based update provides a stable estimate under a limited evaluation budget, as further examined in Appendix~\ref{apd:sensitivity}.

\subsection{Scope}

Our goal in this paper is not to propose a new LLM-MAS architecture or a new graph optimization algorithm, but to study active task selection for communication-structure optimization. Accordingly, our method is designed as a modular layer on top of an existing communication-graph optimization pipeline: it evaluates candidate tasks according to how informative they are for updating the communication structure, while treating the underlying multi-agent system and graph optimizer as black boxes. In this sense, the proposed framework is not inherently tied to a specific number of agents, a particular LLM backbone, or the REINFORCE-based graph optimization method used in our experiments.

More broadly, the proposed framework requires a parameterized communication structure that can be evaluated under different graph parameters, together with a task-dependent scalar or vector performance score. Under these conditions, the proposed EKI-based utility estimator can be applied without differentiating through either the LLM-MAS or the graph optimizer.

Our active task-selection layer is separate from the communication-structure optimization layer: it identifies informative training tasks, while the selected tasks can in principle be used with different graph optimizers. Therefore, the proposed active learning framework is not tied to the REINFORCE-based optimizer adopted in our experiments. Empirical evaluation with other communication-structure optimization methods is left for future work.

Finally, newly obtained information can affect communication-structure optimization at two different levels. First, within each MAS rollout, we keep the communication graph fixed. In multi-step tasks, intermediate environment feedback may change subsequent agent messages or actions, but the graph itself is not re-optimized during the rollout. Second, after a batch of informative tasks is selected and used to optimize the graph, the graph parameters change, and the tasks that are most informative for further optimization may also change. Our task-selection procedure can therefore be reapplied to the updated graph to select the next batch of training tasks. In this work, we focus on fixed-graph execution within each rollout and single-round task selection, leaving within-rollout graph adaptation and multi-round task selection for future study.